\documentclass[12pt]{article}
\pdfoutput=1
\usepackage{amsmath,amssymb}
\usepackage{bbm}
\usepackage{feynmf}
\usepackage{graphicx}
\usepackage{amsbsy}
\newcommand{\3}{\mbox{${\bf \underline{3}}$}}
\newcommand{\s}{\mbox{${\bf \underline{1}}$}}
\newcommand{\spr}{\mbox{${\bf \underline{1}'}$}}
\newcommand{\sppr}{\mbox{${\bf {\underline{1}''}}$}}
\begin{document}
\setlength{\unitlength}{1mm}
\begin{fmffile}{A4warpedpics}
%% Define extension for crosses over lines
\fmfcmd{%
vardef cross_bar (expr p,len,ang)=
((-len/2,0)--(len/2,0))
rotated (ang+angle direction length(p)/2 of p)
shifted point length(p)/2 of p
enddef;
style_def crossed expr p = 
cdraw p;
ccutdraw cross_bar (p, 3mm, 45);
ccutdraw cross_bar (p, 3mm, -45)
enddef;}
%%%
%% Reset size of arrows : default = 4mm and 15
\fmfset{arrow_len}{3.5mm}
\fmfset{arrow_ang}{13}

\begin{titlepage}
\vspace*{0.8cm}
\begin{center}
 {\huge \bf An $A_4$ flavor model for quarks and leptons in warped geometry}
\end{center}
\vskip0.2cm

\begin{center}
 {\bf Avihay Kadosh$^a$ and Elisabetta Pallante$^a$}
\end{center}
\vskip 4pt

\begin{center}
$^a$ {\it Centre for Theoretical Physics, University of Groningen, 9747 AG, Netherlands}

\vspace*{0.1cm}

{\tt a.kadosh@rug.nl, e.pallante@rug.nl }
\end{center}
\vglue 0.3truecm

\begin{abstract}
\vskip 3pt \noindent We propose a spontaneous $A_{4}$ flavor symmetry breaking scheme implemented in a 
warped extra dimensional setup to explain the observed pattern of quark and lepton 
masses and mixings. The main advantages of this choice are the explanation of
fermion mass hierarchies by wave function overlaps, the emergence
of tribimaximal neutrino mixing and zero quark mixing at the
leading order and the absence of tree-level gauge mediated flavor violations. 
Quark mixing is induced by the presence of bulk flavons, which allow for ``cross-brane'' 
interactions and a ``cross-talk'' between the quark and neutrino sectors, realizing the 
spontaneous symmetry breaking pattern $A_4\to {\rm nothing}$ first proposed in 
[X.G.\,He, Y.Y.\,Keum, R.R.\,Volkas, JHEP{0604}, 039 (2006)]. 
We show that the observed quark mixing pattern can be explained in a rather economical way,
 including the CP violating phase, with leading order cross-interactions, while the observed 
difference between the smallest CKM entries $V_{ub}$ and $V_{td}$ must arise from higher 
order corrections. 
Without implementing $P_{LR}$ (or other versions of) custodial symmetry, the bulk mass parameter of the left-handed quarks in this model is constrained by the $Zb\bar{b}$ best fits, still allowing for a Kaluza-Klein scale below 2 TeV.
Finally, we briefly discuss bounds on the Kaluza-Klein scale implied by flavor changing neutral 
current processes in our model 
and show that the residual little CP problem is milder than in flavor anarchic models. 
\end{abstract}

\end{titlepage}

\section{Introduction}

Warped extra dimensions \cite{RS}, which have been proposed as an
alternative solution to the gauge hierarchy problem, also provide a
simple framework in which fermion masses are explained by the
overlap of the fermion and Higgs wave functions in the bulk of the
warped extra dimension \cite{bulkfields}. Having the
zero mode fermions peaked at different points in the fifth
dimension, the exponentially hierarchical masses of quarks and
charged leptons can be obtained with a tiny hierarchy of bulk masses and
 all 5D Yukawa couplings
being of order unity \cite{bulkfermion1, bulkfermion2}.
However, letting the standard model (SM) fermion content
propagate through the bulk generally results in large contributions to
electroweak precision  observables, such as the Peskin-Takeuchi
$S$, $T$  parameters, unless the lowest Kaluza-Klein (KK) mass
scale is unnaturally pushed to values much higher than a TeV. 
To suppress these contributions, more realistic models involving a bulk
custodial symmetry, broken differently at the two branes \cite{Agashe:2003zs} were 
constructed. Alternatively, large brane kinetic terms were introduced
\cite{branekinetic}. In both cases a mass of the first KK excited state as low as $\mathcal{O}$(3 TeV), is 
now allowed by electroweak precision data. 
Another problem arises, this time due to the presence of non degenerate 5D bulk mass
parameters, governing the localization of bulk zero modes. The non degeneracy induces new 
physics (NP) contributions to flavor changing neutral current (FCNC) processes mediated by
 KK excitations of the gauge bosons and fermions, through 
gauge interactions in the fermion kinetic terms and 5D Yukawa interactions.
In the most general case, without imposing any additional flavor symmetry and assuming 
anarchical 5D Yukawa couplings, new physics contributions can already be generated at 
tree level through a KK gauge boson exchange. 
Even if an RS-GIM suppression mechanism \cite{bulkfermion2,rsgim1,Cacciapaglia:2007fw} is at
 work, 
stringent constraints on the KK scale come from the $K^{0}-\overline{K^{0}}$ oscillation
 parameter $\epsilon_{K}$ and the radiative decays $b\to s(d)\gamma$ \cite{Agashe:2004cp}, 
the direct CP violation parameter $\epsilon^\prime/\epsilon_K$ \cite{IsidoriPLB}, and
 especially the neutron electric dipole moment 
\cite{Agashe:2004cp}, where a mass of the first KK state of $\mathcal{O}$(3 TeV) gives rise to a NP 
contribution which is roughly twenty times larger than the current experimental bound -- 
a CP problem in itself, referred to as little CP problem.
Stringent constraints on the KK scale are also present in the lepton sector 
\cite{Kitano:2000wr,lfv,Agashe:2006iy}.
Even in the absence of neutrino masses, severe bounds arise from contributions to FCNC 
processes mediated by tree level KK gauge bosons with anarchical 5D Yukawa couplings 
\cite{Agashe:2006iy}.
It was also recently observed 
\cite{Azatov} that the mixing between fermion zero modes and   
KK modes generally induces a misalignment in the 4D effective 
theory between the SM fermion masses and the Higgs Yukawa couplings. This misalignment 
leads to Higgs mediated flavor changing neutral currents, once the fermion mass matrix is 
diagonalized. In particular, $\epsilon_K$ is 
found \cite{Azatov} to produce stringent combined lower bounds on the KK gluon and the 
standard model 
Higgs mass in flavor anarchic models.  

Additional flavor symmetries in the bulk can in principle allow to partially or fully 
remove these constraints, by forbidding 
or providing a further suppression of tree level FCNCs and one loop 
contributions induced by the presence of KK modes.  
One example that removes or suppresses all tree level contributions 
is the generalization to 5D of minimal flavor violation in the quark sector 
\cite{Fitzpatrick:2007sa,Santiago:2008vq} and in the lepton
sector \cite{Chen:2008qg, Perez:2008ee}.
In these settings, the bulk mass matrices are aligned with the 5D
Yukawa matrices as a result of a bulk $[U(3)]^{6}$ flavor symmetry that is broken in a 
controlled manner.
In \cite{Csaki:2009wc} a shining mechanism is proposed, where the suppression of flavor 
violation in the effective 4D theory on 
the IR brane
is obtained by confining the sources of flavor violation to the UV brane, and communicating 
its effects through gauge bosons of the gauged bulk flavor symmetry.
There, it is also shown that Higgs mediated 
FCNCs are eliminated to leading order, and a lowest KK scale of about 2-3 TeV seems 
to be allowed, rendering the model testable at
collider experiments \cite{collider}.

All above considerations suggest that a candidate for a realistic model of lepton and quark 
masses and mixings in a warped setup 
should possibly be realized with all standard model fields in the bulk, including the Higgs field, a bulk custodial symmetry and an additional flavor symmetry,
 to avoid large new physics contributions and maintain the KK scale of order a 
TeV. 
In \cite{Csaki:2008qq}, see also \cite{delAguila:2010vg}, a bulk $A_{4}$ family symmetry \cite{a4} was
used to explain masses and mixings in the SM lepton
sector. In this setting, the three left-handed lepton doublets form a triplet
of $A_{4}$ to generate tribimaximal (TBM) neutrino mixing
\cite{Harrison:1999cf}, in agreement with the recent global fit in 
\cite{Fogli:2008}. In addition,  tree-level leptonic FCNCs are
absent in this scheme. While the simplest realization of $A_{4}$ well describes the lepton
sector, it does not give rise to a realistic quark sector. In this
paper, we propose a model based on a bulk $A_4$ family symmetry,
implemented in a slightly different setup, in an attempt to
describe both the quark and lepton sectors. In this setup the
scalar fields that transform under non trivial representations of
$A_{4}$, namely two flavon triplets, reside in the bulk.
Consequently, they allow for a complete "cross-talk" \cite{Volkas} between the 
$A_{4}\to Z_{2}$
spontaneous symmetry breaking (SSB) pattern associated with the
heavy neutrino sector - with scalar mediator  peaked towards the UV brane - and the
$A_{4}\to Z_{3}$ SSB pattern associated with the quark and
charged lepton sectors - with scalar mediator peaked towards the IR brane. 
As in previous models based on $A_{4}$, the
three generations of left-handed quarks transform as triplets of $A_4$; this assignment
forbids tree level gauge mediated FCNCs and will allow to obtain realistic masses and 
almost
realistic mixing angles in the quark sector. It will also be instructive to compare this 
pattern to the case of larger realizations 
of the flavor symmetry, like $T'$ \cite{TPrime}, which are usually associated
with a rather richer flavon sector.

An additional feature worth to mention is the constraint on the common left-handed quarks bulk mass parameter implied by the $Zb\bar{b}$ best fits in our model. The numerical significance of such a constraint has been thoroughly investigated in the minimal version of RS models \cite{Casagrande_RSmin}, and it is largely relaxed in models with (extended) $P_{LR}$ custodial symmetry \cite{Agashe:2006at,Casagrande_ECust}.

The paper is organized as follows. In Section~\ref{sec:setup} we
review the basic setup of the model and the various representation
assignments. We then present a RS model with custodial symmetry and a bulk $A_{4}$ family
symmetry, and derive the leading order results for masses and mixings. In Section~\ref{secHO}
we classify all higher order corrections, including cross-talk and cross-brane operators
 for leptons and quarks and parametrize
their effect. Section~\ref{FindCKM} contains our
numerical analysis and results. In Section~\ref{Alignment} we discuss
the vacuum alignment problem and suggest possible solutions, while in Section~\ref{sec:FV} 
we briefly discuss constraints from flavor violating processes on the Kaluza-Klein scale in 
our model. We conclude in Section~\ref{sec:conclusion}.

\section{The model and leading order results}
\label{sec:setup}
We adopt the RS1 framework and thus assume  the bulk of our
model to be a slice of $\mathrm{AdS_5}$, with the extra dimension,
$y$, compactified on an orbifold $S_1/Z_2$ with radius $R$. 
Two 3-branes with opposite tension are 
located at the orbifold fixed points $y=0$, the UV brane, and $y=\pi R$, the IR brane. 
The resulting bulk geometry is described by the metric
\begin{equation}
ds^2=dy^2+e^{-2k|y|}\eta_{\mu\nu}dx^{\mu}dx^{\nu}\label{metric},
\end{equation}
where $k\sim M_{{\rm Pl}}$ is the $\mathrm{AdS}_5$ curvature
scale. The geometric warp factor sequestering the two branes
generates two characteristic scales within this setup, $k$ and $M_{KK}\equiv k\,exp(-k\pi R)$, the latter 
referred to as the KK scale\footnote{Notice that the first KK gauge boson mass is about $2.45\,M_{KK}$}. The electroweak scale
naturally arises for $kR\simeq11$.

All matter fields of our model, fermions and scalars including the Higgs field, live in the 
bulk and we allow for arbitrary $Z_2\times Z_2^\prime$ orbifold boundary conditions, where 
$Z_2$ is the reflection about $y=0$ and $Z_2^\prime$ is the reflection about $y=\pi R$. In 
other words, we allow for discontinuity of the bulk profiles at the orbifold fixed points by
 the presence of non trivial Scherk-Schwarz twists \cite{SStwist}.

The symmetry group in our model is
\begin{equation}
\mathop{G} =G^{cust}_{SM} \times A_4 \times Z_2 =\mathop{\rm
SU}(3)_{c} \times \mathop{\rm SU}(2)_{L}\times \mathop{\rm
SU}(2)_{R} \times \mathop{\rm {}U}(1)_{B-L} \times A_4\times
Z_{2}\, .
\end{equation}
The bulk gauge symmetry $G^{cust}_{SM}$ is augmented by an $A_4$ flavor symmetry
plus an auxiliary $Z_{2}$, whose nature and role will be
explained below. In addition, the electroweak gauge group is extended to 
$\mathop{\rm SU}(2)_{L}\times 
\mathop{\rm SU}(2)_{R} \times \mathop{\rm {}U}(1)_{B-L} $ to incorporate custodial symmetry 
\cite{Agashe:2003zs}, and thus protect electroweak
precision measurements with the lightest Kaluza-Klein mass
being as light as ${\cal O}$(4TeV). This symmetry is broken down to
the SM group $\mathop{\rm SU}(2)_{L}\times \mathop{\rm {}U}(1)_{Y} $ on the UV brane, 
and down to $\mathop{\rm SU}(2)_{D}\times \mathop{\rm {}U}(1)_{B-L}$ on the IR brane. 

Both breaking patterns can be realized by orbifold boundary conditions on the gauge fields 
under $Z_2\times Z_2^\prime$ as in \cite{Agashe:2003zs}. In particular, the complete 
UV breaking pattern is achieved via an $SU(2)_R$ doublet
$(1,2)_{1/2}$ or a triplet scalar VEV, while a bidoublet $(2,2)_{0}$ Higgs VEV induces 
the IR breaking.
These fields can either be decoupled by taking their infinite mass limit as in higgsless
 models, or be dynamical and used to generate masses of quarks and leptons, as it is true in 
our case for the Higgs field. Notice also that, in our case, the bidoublet Higgs field 
lives in the bulk and it is peaked towards the IR brane.

We introduce two scalar flavons $\Phi$ and $\chi$ and a Higgs field transforming under 
$G^{cust}_{SM} \times A_4$ as 
\begin{equation}
\Phi \sim \left( 1,1,1,0 \right) \left( \3 \right),\quad \chi \sim
\left( 1,1,1,0 \right) \left( \3 \right),\quad H\left(1,2,2,0
\right) \left( \s \right)\, .
\end{equation}
The three families of quarks and leptons are assigned to the
following representations:
\begin{equation}
\begin{array}{c}
Q_L \sim \left( 3,2,1,\frac{1}{3} \right) \left( \3 \right) \\
\\
u_R \oplus u'_R \oplus u''_R \sim \left( 3,1,2,\frac{1}{3} \right)\left(\s \oplus 
\spr \oplus \sppr \right) \\
\\
d_R \oplus d'_R \oplus d''_R \sim \left( 3,1,2,\frac{1}{3}
\right)\left(\s \oplus \spr \oplus \sppr \right)
\end{array}
\quad\
\begin{array}{c}
\ell_L \sim \left( 1,2,1,-1 \right) \left( \3 \right) \\
\\
\nu_R \sim \left( 1,1,2,0 \right)\left( \3 \right) \\
\\
e_R \oplus e'_R \oplus e''_R \sim \left( 1,1,2,-1 \right)\left(\s
\oplus \spr \oplus \sppr \right)\, ,
\end{array}
\end{equation}
where the $A_4$ notation is explained in the appendix, and the
$\mathop{\rm G}^{cust}_{SM}$ notation is standard. The $Z_{2}$
assignments will be specified later. Models with similar A$_4$
assignments for leptons and scalar fields have been considered
before \cite{Volkas,a4},  thus many of the leading order properties are shared with our
 model.
Notice that the right-handed neutrinos are assigned
to a $\3$ of $A_4$, whereas the right-handed charged-fermions are each
given a $\s \oplus \spr \oplus \sppr$ structure. It is also
important to notice that we have a separate $\mathop{\rm SU}(2)_R$
doublet for each right handed fermion.

Bulk fermions are assigned specific parities under $Z_2\times Z_2^\prime$, so that 
zero modes  will provide the standard model particle content. As implied by the invariance 
of the 5D action, the bulk fermion field $\Psi$ transforms as 
$\Psi (y) = Z\gamma_5\Psi (-y)$ and $\Psi (\pi + y) = Z^\prime\gamma_5\Psi (\pi -y)$, 
with $Z, \, Z^\prime =\pm 1$,
under $Z_2$ and $Z_2^\prime$, respectively.  
The 5D fermion corresponds to two Weyl spinors of opposite chirality in 4D
\begin{equation}
\Psi = \left( \begin{array}{c} \xi \\ \bar{\psi}
\end{array}�\right)
\end{equation}
and with opposite $Z_2$ and $Z_2^\prime$ parities. Hence, if $\xi$ has parities 
$(Z,Z^\prime )=(+,\,+)$, $\bar{\psi}$ has parities $(-,\, -)$. Even parity at $y=0$ or 
$y=\pi R$ implies Neumann-like boundary conditions on the bulk fermion profile, while odd 
parity implies Dirichlet boundary conditions. It follows that a massless zero-mode only 
exists for a 
$(+,\, +)$ Weyl spinor. 

The bulk profile of the would-be zero mode is shaped by the fermionic bulk mass term with 
mass $m=\epsilon (y) c k$, and $\epsilon (y)$ is the sign function.
For $c>1/2$ (resp. $c<1/2$), the zero mode is exponentially localized on the UV
(IR) brane. Finally, it is important to notice that since $\mathop{\rm SU}(2)_R$ is broken 
on the UV brane, the two components of each $\mathop{\rm SU}(2)_R$ doublet must have
 opposite $Z_2$ parities. To get a massless zero mode from both components, we thus need to 
double the number of doublets \cite{Agashe:2003zs}, and we do this for quarks and leptons. 
Taking into account the above considerations, we assign the following boundary conditions to
 leptons,
in the absence of localized mass terms:
\begin{equation}
\begin{array}{ccc}
\ell_L=\left(\begin{array}{cc} L & [+,+]\end{array}\right) &\,\,
e_R, \mu_R, \tau_R=\left(\begin{array}{cc}
 \tilde{\nu}_{e,\mu,\tau} & [+,-]\\e_R,\mu_R,\tau_R & [-,-]\end{array}\right) & \,\,
\nu_R=\left(\begin{array}{cc}
\nu_R & [-,-]\\\tilde{\ell} & [+,-]\end{array}\right)
\end{array},
\end{equation}
where the parities $(Z,\, Z^\prime )$ are given for the upper Weyl spinor $\xi$, while 
$\bar{\psi}$ has the opposite conditions.
Hence, there is a left handed zero mode for each left
handed doublet in $\ell_L$, $\ell = e,\,\mu ,\,\tau$, and a single right handed zero mode
in $e_R$, $\mu_R$, $\tau_R$ and $\nu_R$ each.
These fields have bulk masses, in units of the AdS curvature,
given by $c_{{\ell}\, L}, c_{{\ell}\, R}$ and $c_{\nu R}$, with  $\ell = e,\,\mu ,\,\tau$, 
and we work in the basis where they are real and diagonal. 
An important restriction is due to the A$_4 (\times$Z$_2)$ bulk global symmetry: since the 
three left handed lepton
doublets are unified into a triplet of A$_4$, they will share
one common $c$ parameter which we label $c_L^{\ell}$.

Analogously, the boundary conditions for the quarks are chosen to be:
\begin{equation}
\begin{array}{ccc}
Q_L=\left(\begin{array}{cc} Q_L & [+,+]\end{array}\right) & u_R,
c_R, t_R=\left(\begin{array}{cc} u_R, c_R, t_R & [-,-]
\\\tilde{d}, \tilde{s}, \tilde{b}&[+,-]\end{array}\right) & d_R,
s_R, b_R=\left(\begin{array}{cc} \tilde{u}, \tilde{c}, \tilde{t} &
[+,-]\\d_R, s_R, b_R& [-,-]\end{array}\right)
\end{array}.
\end{equation}
In this way we have a left handed massless zero mode for the three left handed
doublets in $Q_L$ and a single right handed zero mode in $u_R$,
$c_R$, $t_R$, $d_R$, $s_R$ and $b_R$ each. Again, the three left handed
quark doublets, being  assigned to a triplet of A$_4$, share one common
bulk mass parameter, $c_L^q$. The right handed quarks  are
assigned to distinct one dimensional representations of A$_4$,
hence there are 6 different bulk mass parameters entering their zero mode profiles, 
$c_i^{u,d}$ and $i=1,2,3$. 

The $G$ invariant 5D Yukawa lagrangian at leading order reads
\begin{eqnarray}
{\cal L}_{{\rm Yuk(5D)}}  = & \Lambda_{5D}^{-2} [&y_u(
\overline{Q}_L \Phi )_{\s}\,H\, u_R + y_u' ( \overline{Q}_L \Phi
)_{\spr}\,H\, u''_R
+ y_u'' ( \overline{Q}_L \Phi )_{\sppr}\,H\, u'_R  \nonumber\\
& + &  y_d ( \overline{Q}_L \Phi)_{\s}\,\tilde{H}\, d_R + y_d' (
\overline{Q}_L \Phi )_{\spr}\,\tilde{H}\, d''_R +y_d'' (
\overline{Q}_L \Phi )_{\sppr}\,\tilde{H}\, d'_R   \nonumber\\
& + & y_e ( \overline{\ell}_L \Phi)_{\s}\,\tilde{H}\, e_R + y'_e (
\overline{\ell}_L \Phi )_{\spr}\,\tilde{H}\,e''_R + y''_e (
\overline{\ell}_L \Phi )_{\sppr}\,\tilde{H}\, e'_R \,\,] \nonumber\\
& + & \Lambda_{5D}^{-1/2}\left[y_{\nu}^{D}( \overline{\ell}_L
\nu_R )_{\s}\,H\, + y_\chi[\overline{\nu}_R (\nu_R)^c ]_{\3s}
\cdot \chi\right] + M [ \overline{\nu}_R (\nu_R)^c ]_{\s} + h.c.\, ,
\label{eq:Yuk}
\end{eqnarray}
where $\tilde{H} \equiv i\tau_{{2}_L} H^*$, all fields propagate in the bulk and 
$\Lambda_{5D}$ is naturally of order 
$M_{{\rm Pl}}$. 
The Higgs and $\Phi$ fields are chosen to be peaked towards the IR
brane at $y=\pi R$, while the field $\chi$ is peaked towards the UV
brane at $y=0$, for phenomenological reasons. The
lagrangian in Eq.~(\ref{eq:Yuk}) has a relatively simple
structure. Each charged fermion sector $u$, $d$, $e$ has three
independent Yukawa terms, all involving the $A_4$ triplet field
$\Phi$  and the Higgs field. By construction, the neutrino Dirac
term is governed by a single coupling constant $y_\nu^D$
and involves the Higgs only, while the right-handed Majorana
sector contains one bare Majorana mass $M$, and a single Yukawa
coupling $y_\chi$ to the electroweak singlet and $A_4$ triplet
$\chi$. %\footnote{Note that the $\3a$
%product of $\overline{\nu}_R $ and $(\nu_{R})^c $ identically
%vanishes.}
In total there are only twelve, a priori complex, Yukawa parameters
to describe masses and mixings of nine Dirac and six Majorana
fermions. All of these parameters will be  taken to be universal
and of $\mathcal{O}(1)$ in the construction of Sec.~\ref{FindCKM}.

The lagrangian in Eq.~(\ref{eq:Yuk}) respects an additional
$Z_{2}$ symmetry, under which $Q_L$, $\ell_L$, $\nu_{R}$ and
$\Phi$ are odd, while all other fields are even. This
non-flavor symmetry ensures that the $\mathop{\rm G}_{\rm{SM}}^{\rm{cust}}\times A_4$ 
invariant term $\overline{\ell}_L \Phi H\nu_R$ is absent from the Lagrangian.

All the SM fields are identified with the 4D components of the zero modes in the Kaluza-Klein
 decomposition of the bulk fields. 
Masses and mixings for leptons and quarks are induced by Eq.~(\ref{eq:Yuk}) once the flavons
 $\Phi ,\, \chi$ and the Higgs have acquired a VEV. The VEVs of $\Phi$ and $\chi$ will be 
responsible for  
providing two distinct patterns of spontaneous symmetry breaking of 
A$_4$.
The VEV profiles for the scalar fields in our model are solutions of the bulk equations of
motion with almost vanishing bulk mass for stabilization purposes \cite{WiseScalar}, an IR 
localized quartic double well 
potential for $\Phi$ and the Higgs, and a similar term for $\chi$ on the UV
brane. To leading order in $exp(-2\pi kR)$, they read 
\begin{equation}
 \Phi_a  (y)= v_ae^{4(k|y|-\pi kR)}\qquad
H(y) = H_{0}e^{4(k|y|-\pi kR)}\qquad
\chi_a(y)=\chi_a(1-e^{4(k|y|-\pi
kR)})\,,\label{profiles}
\end{equation}
where $a=1, 2, 3$ denotes the $A_4$ component.
The size of the SM fermion masses is thus determined
by the amount of wavefunction overlap of two zero modes of opposite
chirality, corresponding to the same Dirac fermion in 4D,
together with the VEV profiles of the corresponding scalar fields.
This holds as far as the zero mode approximation (ZMA) is a good description of the 4D 
reduction of our model. 
In general, the presence of kinetic and potential boundary terms
for the scalar fields will, after SSB of A$_4$
and the electroweak symmetry,  lead to boundary conditions that
mix all levels of KK fermions. The light modes are however
rather insensitive to the presence of these boundary terms and can
be treated as a small perturbation. Thus, to leading order, the
low energy mass spectrum can be obtained by using the zero mode
profiles of all bulk fields. 
The accuracy of the ZMA in each specific case depends on the lightness of the
lowest lying KK states. Since the largest mass present  in our
model is that of the $t$ quark, $m_t=171.3$ GeV, the ZMA turns out
to be as accurate as $m_t\,M_{KK}\simeq 0.03$ for the masses
of the zero mode fermions.

Writing out the charged-fermion $f=u,d,e$ Yukawa invariants of Eq.~(\ref{eq:Yuk}), and 
following the rules %\ref{eq:33tos}-\ref{eq:33tosppr}
in the appendix, one finds that each of the three mass matrices
has the form
\begin{equation}
M_f = \int_{-\pi R}^{\pi R}~dy\, \sqrt{-g}\,\frac{H(y)}{\Lambda_{5D}^{2}} 
\left( \begin{array}{ccc}
\overline{f}_{1L},\overline{f}_{2L},\overline{f}_{3L}
\end{array} \right)
\left( \begin{array}{ccc}
y \Phi_1 & \ \ y' \Phi_1 & \ \ y'' \Phi_1 \\
y \Phi_2 & \ \ \omega y' \Phi_2 & \ \ \omega^2 y'' \Phi_2 \\
y \Phi_3 & \ \ \omega^2 y' \Phi_3 & \ \ \omega y'' \Phi_3
\end{array} \right)
\left( \begin{array}{c} f_R \\ f''_R \\ f'_R
\end{array} \right) + h.c.
\end{equation}
where the scalar profiles $H(y)$ and $\Phi_a (y)$ are from Eq.~(\ref{profiles}), 
the metric factor is
 $\sqrt{-g}=\exp(-4k|y|)$, the Yukawas $y,y',y''$ have a suppressed subscript $f$, 
and $f_{L,R}(y)$ are the fermion bulk profiles. 
The numerical subscripts $1,2,3$ denote $A_4$ components, as in
the appendix. For the special VEV pattern of $\Phi$
\begin{equation}
v_1 = v_2 = v_3 \equiv \Phi_0 \label{eq:C3vac},
\end{equation}
we define $v\equiv H_0\Phi_0/\Lambda_{5D}^{2}$, and each of the
above mass matrices translates into the effective 4D mass
matrix
\begin{equation}
M_f = U(\omega) \left( \begin{array}{ccc} \sqrt{3}\tilde{y}_f v & 0 & 0 \\
0 & \sqrt{3} \tilde{y}'_f v & 0 \\ 0 & 0 & \sqrt{3} \tilde{y}''_f
v
\end{array} \right)\,\,\,\,\,\,\,\, \tilde{y}_{f}=y_{f}\int^{\pi R}_{-\pi\textwidth 6.5in \oddsidemargin 0in \evensidemargin 0in \textheight
8.5in  \topmargin -0.2in
R} \frac{dy}{2\pi R} F(c_{L_{f}}, c_{R_{f}})e^{8k|y|-8k\pi R} \, , \label{eq:Mf}
\end{equation}
where we conveniently introduced the fermion overlap function
\begin{equation}
F(c_{L_{i}}, c_{R_{j}}) \equiv \sqrt{-g}f^{(0)}_{L_i}(y) f^{(0)}_{R_j}(y)= k\pi R \sqrt{ \frac{ (1 - 2 c_{L_{i}})
(1 - 2 c_{R_{j}})}{ (e^{(1 - 2 c_{L_{i}}) \pi k R} - 1)(e^{(1 - 2
c_{R_{j}}) \pi k R} - 1)} } e^{(- c_{L_{i}} - c_{R_{j}})  k |y|} \
, \label{eqn:ffactor}
\end{equation}
product of the 5D profiles for the zero modes
of the fermion fields, $f^{(0)}_{L_i}$ and $f^{(0)}_{R_j}$, with
$f=u,d,e$ and the factor $\sqrt{-g}$ included. 
This shows that the left-diagonalization matrices $V_L^{u,d,e}$ for the up-quark,
 down-quark and charged lepton 
sectors, respectively,
are identical and equal to the unitary trimaximal mixing
matrix~\cite{Volkas},
\begin{equation}
V_L^{u,d,e}=U(\omega) = \frac{1}{\sqrt{3}} \left(
\begin{array}{ccc} 1 & 1 & 1
\\ 1 & \omega & \omega^2 \\ 1 & \omega^2 & \omega
\end{array} \right).
\end{equation}
This diagonalization property of mass matrices is referred
to as ``form diagonalizability''  \cite{lowvolkas}; in this case the mixing angles are
 independent of the mass eigenvalues.
 One can easily see that, at this order and due to the above $A_4$ assignments, the
$\Phi$ VEV of Eq.~(\ref{eq:C3vac}) forces the CKM
matrix to be the identity:
\begin{equation}
V_{CKM} = V_L^{u\dagger} V_L^d = U(\omega)^{\dagger} U(\omega) = 1\, .
\end{equation}
It also induces the breaking pattern
\begin{equation}
A_4 \to Z_3 \cong C_3 = \{1,c,a\},
\end{equation}
where $\cong$ denotes ``isomorphism'', see the appendix. The remnant  $Z_3$ flavor
subgroup cyclically permutes the three $A_4$ triplet basis states
with no change of signs, see Eq.~(\ref{eq:ca}) and \cite{Volkas}. The
$\spr$ and $\sppr$ representations transform under this subgroup
in the same way they do under the full flavor group, $A_4$. We
will show below that, if the remnant $Z_3$ symmetry remains
unbroken,  the CKM matrix will remain trivial to all orders.
A further breaking is thus needed in order to produce deviations of the CKM matrix from
 unity.
In \cite{Volkas} it was first suggested that such  small deviations 
 can be generated by higher-order effects, able to induce a
relatively weak subsequent breaking of the residual $Z_3$ flavor
symmetry, i.e. $Z_3\to nothing$.

Our main goal is to implement this idea in our setup 
to obtain a realistic CKM matrix, without spoiling the results
that will be obtained below for the neutrino and charged lepton
sector. It is for this purpose that we choose to use bulk flavons.
The non vanishing overlap between the profiles of $\Phi$ and
$\chi$ allows for the presence of higher dimensional operators, which communicate
the symmetry breaking pattern associated with $\langle\chi\rangle$ to the quark
sector, as explained in section~\ref{Crosstalk1}. On the
other hand, having these profiles exponentially localized on
different branes, combined with the internal symmetries of the
model, strongly suppresses the mixed interaction terms in
the scalar potential $V(\Phi ,\chi)$, thus allowing to approximately preserve the
original alignment between the VEVs of these fields, as discussed in
Sec.~\ref{Alignment}.

\subsection{Leading order results in the neutrino sector.}

 This is an immediate generalization to our model of the 4D derivation in 
\cite{Volkas} and the 5D warped model with brane localized scalars on an interval in 
\cite{Csaki:2008qq}. 
The neutrino Dirac mass matrix is  derived from the Yukawa
term $\overline{\ell}_L H\nu_R$ in Eq.~(\ref{eq:Yuk}).  Using Eq.~(\ref{eq:33tos}),
 the 4D Dirac mass matrix turns out to be proportional to the identity matrix
\begin{equation}
\label{eq:MD}
M_\nu^D = H_0\tilde{y}_\nu^{D} \, \mathbbm{1}=
\left(\frac{H_{0}y_{\nu}^{D}}{\Lambda_{5D}^{1/2}}\int^{\pi
R}_{-\pi R} \frac{dy}{2\pi R}\,  e^{4k(|y|-\pi R)}F(c_L^{\ell}, c_{\nu _{R}})\
\right)\cdot\mathbbm{1} \equiv m_\nu^D\, \bf{\mathbbm{1}}\,\, .
\end{equation}
In this equation
$\tilde{y}_{\nu}^D$ is the effective 4D coupling and $y_{\nu}^D$
is  dimensionless. The right-handed neutrino bare Majorana mass
matrix is similarly trivial, being 
$M_{\nu}^{M}=M\int dy\, F(c_{\nu _{R}}, c_{\nu _{R}})
\cdot\mathbbm{1}\equiv\tilde{M}\,\mathbbm{1}$. It is the  coupling to
the flavon $\chi$ that induces a non trivial pattern in the neutrino mass matrix, with 
contribution
\begin{equation}
\label{eq:Mchi}
 M_{\nu}^{\chi}=\tilde{y}_{\chi}
\left( \begin{array}{ccc} 0 & \chi_3 & \chi_2 \\ \chi_3 & 0 &
\chi_1 \\ \chi_2 & \chi_1 & 0
\end{array} \right)\, ,
 \qquad
\tilde{y}_\chi=\frac{y_{\chi}}{\Lambda_{5D}^{1/2}}\int^{\pi
R}_{-\pi R}\;\frac{dy}{2\pi R}\, (1-e^{4k(|y|-\pi R)})F( c_{\nu _{R}},  c_{\nu _{R}})\, ,
\end{equation}
with $\chi_a, \, a=1,2,3$ the VEV components of Eq.~(\ref{profiles}).  We
now follow \cite{a4,Volkas} and assume the breaking pattern $A_4 \to Z_2 = \{1,r_2\}$
induced by the choice of the $\chi$ VEV
\begin{equation}
\chi_1 = \chi_3 = 0,\quad
\chi_2 \equiv \chi_0 \neq 0\, , \label{eq:Z2vac}
\end{equation}
so that the full  $6 \times 6$ neutrino mass matrix in 4D becomes
\begin{equation}
M_{\nu}^{total}=\left( \begin{array}{cccccc}
0 & 0 & 0 & m_{\nu}^D & 0 & 0 \\
0 & 0 & 0 & 0 & m_{\nu}^D & 0 \\
0 & 0 & 0 & 0 & 0 & m_{\nu}^D \\
m_{\nu}^D & 0 & 0 & \tilde{M} & 0 & M_\chi \\
0 & m_{\nu}^D & 0 & 0 & \tilde{M} & 0 \\
0 & 0 & m_{\nu}^D & M_\chi & 0 & \tilde{M}
\end{array} \right),
\end{equation}
where $M_\chi \equiv \tilde{y}_\chi \chi_0$, and $\tilde{M}$
and $M_\chi$ are in general complex. In the see-saw limit
$|\tilde{M}|, |M_\chi| \gg m_{\nu}^D$, and the effective $3 \times 3$
light neutrino mass matrix is thus
\begin{equation}
M_L^{\nu} = - M_\nu^D \left ( M_{\nu}^M +M_\nu^\chi\right )^{-1} (M_\nu^D)^T = -
\frac{(m_\nu^D)^2}{\tilde{M}} \left( \begin{array}{ccc}
\frac{\tilde{M}^2}{\tilde{M}^2-M^2_\chi}  & 0 & - 
\frac{\tilde{M} M_\chi}{\tilde{M}^2-M^2_{\chi}} \\
0 & 1 & 0 \\
- \frac{\tilde{M} M_\chi}{\tilde{M}^2-M^2_{\chi}} & 0 &
\frac{\tilde{M}^2}{\tilde{M}^2-M^2_\chi}
\end{array} \right),
\label{eq:MLbare}
\end{equation}
whose diagonalization matrix is
\begin{equation}
V_L^{\nu} = \frac{1}{\sqrt{2}} \left( \begin{array}{ccc} 1 & 0 &
-1 \\ 0 & \sqrt{2} & 0 \\ 1 & 0 & 1
\end{array} \right).
\end{equation}
The MNSP matrix, at this order, is then
\begin{equation}
V_{MNSP} = V_L^{e\dagger} V_L^{\nu} = U(\omega)^{\dagger}
V_L^{\nu}
 = \left( \begin{array}{ccc}
\frac{2}{\sqrt{6}} & \frac{1}{\sqrt{3}} & 0 \\
-\frac{\omega^2}{\sqrt{6}} & \frac{\omega^2}{\sqrt{3}} & -\frac{e^{-i\pi/6}}{\sqrt{2}} \\
-\frac{\omega}{\sqrt{6}} & \frac{\omega}{\sqrt{3}} &
\frac{e^{-5i\pi/6}}{\sqrt{2}}
\end{array} \right)\,,
\end{equation}
which is tribimaximal up to phases and in good agreement with
the neutrino oscillation data \cite{Fogli:2008}, as already concluded in 
\cite{Volkas,a4he}. Notice that the
Jarlskog invariant is vanishing despite the presence of these
phases, and consequently CP violation is absent at this order.

The bulk scalar fields $\chi$ and $\Phi$, 
which are in charge of the symmetry breaking pattern $A_4\to Z_2$ in the neutrino sector and
$A_4\to Z_3$ in the charged-fermion sector, respectively are peaked on different
branes. Thus, the two distinct flavor symmetry breaking patterns
 will be approximately, but not fully sequestered
from one another, due to the bulk nature of these fields. Thus, while the leading order 
lagrangian in Eq.~(\ref{eq:Yuk}) does not allow for a talking between the two sectors, higher dimensional operators will ensure the complete breaking of the $A_4$ flavor symmetry 
through the overlap of bulk scalar fields. It is these effects that we are interested in, in order to
account for a realistic CKM matrix.

Higher order effects were naturally divided into two classes \cite{Volkas}, those 
 which preserve the flavor subgroups ($Z_2$ or $Z_3$) of each sector,  call them 
``higher-order'', and those that involve interactions between the
two sectors, ``cross-talk''. The former preserve
$Z_3$ in the quark and charged lepton sectors, and $Z_2$ in the
neutrino sector. The latter communicate $Z_3$ breaking to the charged sector via $\chi$, 
and $Z_2$ breaking to the neutrino sector via $\Phi$. 
In our context, a further way to isolate the dominant contributions amongst all higher order 
terms is to distinguish between brane localized interactions, UV or IR, and ``cross-brane''
interactions induced by the overlap of the bulk profiles of $\Phi$ and $\chi$. 
 We will first show the pattern of all dominant higher order corrections 
to the CKM and MNSP matrices, producing deviations from the
trivial and tribimaximal forms, respectively. We will then estimate their structure and size 
in our model and compare with existing results. 
We are mainly interested in the quark sector.
The reason is that within our setup an almost  realistic structure for the
CKM matrix can emerge with a fairly restricted choice of the
parameters involved. 
%In addition, protecting the vacuum alignment and guaranteeing the 
%stability of the leading order results can be achieved with  no further
%field content and internal symmetries, as discussed in section  \ref{Alignment}.

\section{Higher order and cross-talk corrections}
\label{secHO}

We first consider charged fermions, to immediately show that higher order 
contributions in the $Z_3$-preserving sector, i.e. no cross-talk, leave unchanged the 
leading 
order result for mass and mixing matrices, in particular $V_L^\ell =U^\dagger (\omega )$ 
to all orders for charged leptons. 

The leading order contribution comes from the operator in Eq.~(\ref{eq:Yuk}) of the form  
$\bar{f}_L \Phi Hf_R$. The only higher order corrections allowed in the $Z_3$-preserving 
sector come from the operators of the type $\bar{f}_L \Phi^{n} Hf_R$.
 After breaking of $A_4$, and since the VEV of $\Phi$ is $Z_3$ symmetric, 
$\Phi^2$ transforms as $1+\Phi$.
Given that only the $A_4$ triplet part of $\Phi^n$ will contribute, it is clear that 
the corrections to masses and mixings of the above operators to all orders will be identical 
in structure to the leading order ones. 
This shows that the $Z_3$ symmetry in the charged fermion sector will ensure 
$V_L^{\ell ,u,d} = U^\dagger (\omega )$ and thus prevent quark mixing, 
so that the only way to allow for a non trivial CKM matrix is to further break $Z_3$. 
We  assume $V_L^\ell = U^\dagger (\omega )$ 
in the rest of this section, and postpone to sections \ref{Crosstalk1} and 
\ref{CrossLep} the analysis of the suppressed cross-talk $Z_3$ violating contributions.

\subsection{Cross-talk and cross-brane effects in the neutrino sector}
%%%%%%%%%%%%%%%%%%%%%%%%%%%%%%%%%%%%%%

We now identify all the $A_4$ symmetric higher dimensional operators contributing to the 
neutrino sector in our model. These are obtained by additional insertions of the fields $\chi$ and $\Phi$ into the leading order terms for neutrinos in Eq.~(\ref{eq:Yuk}).
Cross-talk is induced by all contributions involving $\Phi$. 
We can already anticipate a pattern in the corrections. Given that the VEV of $\chi$ is 
$Z_2$ preserving, $\chi^3$ transforms effectively as $\chi$ under $A_4$. Hence, the 
contributions of $\chi^m$ operators to the 
  $(1,2)$, $(2,1)$, $(2,3)$ and $(3,2)$ entries of
both the neutrino Dirac and right-handed Majorana mass matrices
will be zero to all orders.
Analogously, since $\Phi^2$ transforms as $1+\Phi$, the contributions of $\Phi^{2n}$ 
operators will be absorbed in the leading order contributions, or effectively 
amount to the operator with one $\Phi$ insertion.
In general, the size of all higher order contributions will be suppressed with respect to
 the leading order by powers of the 
relevant scales $(VEV)^n/\Lambda_{5D}^n$ and by the amount of overlap of bulk profiles; 
recall that $\chi$ is UV peaked, while $H$ and $\Phi$ are IR peaked. Hence, a strong 
suppression is induced by the small overlap in the interference of $\chi -\Phi$ and 
$\chi - H$. 
The complete higher order contribution to the neutrino Dirac lagrangian can be written as 
follows
\begin{equation}
\Delta{\cal L}_{\nu}^{D}=
\frac{1}{\Lambda_{5D}^{(6n+3m+1)/2}}\left[\left.
(\overline{\ell}_{L}\chi^{m}H\nu_{R})_{m\geq 1} \right|_{\rm h.o.} +
\left.\left( 
(\overline{\ell}_{L}\Phi^{2n}H\nu_{R})_{n\geq 1} +
(\overline{\ell}_{L}\Phi^{2n}\chi^{m}H\nu_{R})_{m,n\geq 1}
\right)
\right|_{\rm cross-talk}\right]\, , 
\label{DiracCorr.}
\end{equation}
where we separated the cross-talk terms from the rest. Notice that odd powers of $\Phi$ are 
forbidden by the additional $Z_2$ in $G$. 
It is immediate to obtain the following textures for the above corrections to the neutrino 
Dirac mass matrix to all orders:
\begin{equation}
\frac{M_{\nu}^D}{\tilde{y}_{\nu}^{D}\,H_0}=\left. 
\left(
\begin{array}{ccc} 1 & 0 & 0 \\ 0 & 1 & 0 \\ 0 & 0 & 1
\end{array} \right) \right|_{\rm l.o.} +
\left. \left( \begin{array}{ccc}
\epsilon_{11}^{\chi} & 0 & \epsilon_{13}^{\chi} \\ 
0 & \epsilon_{22}^{\chi} & 0 \\ 
\epsilon_{13}^{\chi} & 0 & \epsilon_{11}^{\chi *}
\end{array} \right) \right|_{\rm h.o.} + 
\left. \left( \begin{array}{ccc}
\epsilon_{11}^{\Phi}+\hat{\epsilon}_{11} & \epsilon_{2}^{\Phi}+ \hat{\epsilon}_{1}& 
\epsilon_{3}^{\Phi}  \\
\epsilon_{3}^{\Phi}+\hat{\epsilon}_{2} & \epsilon_{11}^{\Phi}+\hat{\epsilon}_{22} & 
\epsilon_{2}^{\Phi}+ \hat{\epsilon}_{2} \\
\epsilon_{2}^{\Phi} & \epsilon_{3}^{\Phi}+\hat{\epsilon}_{1} & 
\epsilon_{11}^{\Phi}+\hat{\epsilon}_{11}^{\,*}
\end{array} \right) \right|_{\rm cross-talk}
 \, ,
\label{DiracTextures}
\end{equation}
where $\epsilon_{11,22}^{\chi}\sim{\cal O}(\chi_0^2/\Lambda_{5D}^{3})$, 
$\epsilon_{13}^{\chi}\sim{\cal O}(\chi_0/\Lambda_{5D}^{3/2})$, 
$\epsilon_{i}^{\Phi},\epsilon_{ii}^{\Phi} \sim{\cal O}(\Phi_0^2/\Lambda_{5D}^{3})$ and  
$\hat{\epsilon}_{i}, \hat{\epsilon}_{ii}\sim{\cal O}(\Phi_{0}^{2}\chi_0/\Lambda_{5D}^{9/2})$.
Notice that in the above expression the coefficient
$\epsilon_{11}^{\Phi}$ can be absorbed into a redefinition of
$\tilde{y}_{\nu}^{D}$, and similarly the coefficients
$\hat{\epsilon}_{11}$ and $\hat{\epsilon}_{22}$ can be absorbed in
$\epsilon_{11}^{\chi}$ and $\epsilon_{22}^{\chi}$.
The Majorana mass matrix is corrected by the operators
\begin{equation} \Delta{\cal
L}_{\nu}^{M}=\frac{1}{\Lambda_{5D}^{(3m+6n-2)/2}}\left[\left.
\left(\chi^{m}\overline{\nu}_{R}(\nu_{R})^{c} \right)_{m\geq 2} \right|_{\rm
h.o.} +
\left.\left(\Phi^{2n}\chi^{m}\overline{\nu}_{R}(\nu_{R})^{c}\right)_{m\geq 0,n\geq 1}
\right|_{\rm cross-talk}\right]\, . 
\label{MajoranaCorr.}
\end{equation}
Using again the transformation properties of $\Phi^{2n}$ and
$\chi^{m}$ we obtain the following texture for the Majorana mass matrix,
\begin{equation}
M_{\nu}^M = \left. \left( \begin{array}{ccc} \tilde{M} & 0 & M_{\chi} \\ 0 & \tilde{M} & 0 \\
M_{\chi }& 0 & \tilde{M}
\end{array} \right) \right|_{\rm l.o.} +
\left. \left( \begin{array}{ccc} \epsilon'_{11} & 0 & \epsilon'_{13} \\
0 & \epsilon'_{22} & 0 \\ \epsilon'_{13} & 0 & \epsilon^{'*}_{11}
\end{array} \right) \right|_{\rm h.o.} + \left. \left( \begin{array}{ccc} 0 & \tilde{\epsilon}_{1}+ \epsilon_{1}&
\epsilon_{2} \\
 \tilde{\epsilon}_{2}+ \epsilon_{2} &0 & \tilde{\epsilon}_{2}+ \epsilon_{1} \\\epsilon_{1} &
\tilde{\epsilon}_{1}+ \epsilon_{2} & 0
\end{array} \right) \right|_{\rm cross-talk},
\label{Majtexture}
\end{equation}
where  $\epsilon_{ij}^{'}\sim{\cal O}(M_\chi \chi_0^2/\Lambda_{5D}^{3})$, 
$\tilde{\epsilon}_{i}\sim{\cal
O}(M_\chi\Phi_0^2\chi_0/\Lambda_{5D}^{9/2})$ and
$\epsilon_{i}\sim{\cal O}(M_\chi\Phi_0^2/\Lambda_{5D}^{3})$. 
Notice that $\epsilon_{13}^{'}$ can be absorbed into a redefinition 
of $M_\chi$, while for simplicity we have neglected diagonal contributions that 
redefine $\tilde{M}$.  

If we neglect cross-brane interactions, induced by the overlap of $\Phi$, $\chi$ and $H$ bulk 
profiles, the corrected light neutrino mass matrix, $M_L^\nu$, once rotated to the basis of
diagonal charged leptons with $U^{\dag}(\omega)$, and assuming all real input parameters,
will be to any order of the following form:
\begin{equation}
M_L^{\nu} \to {\bf\tilde{M}_L^{\nu}} = \left. \left(
\begin{array}{ccc} \delta_{1}
& \delta_{2} & \delta_{2}^{*} \\ \delta_2 & \delta_{4} & \delta_3 \\
\delta_{2}^{*} & \delta_{3} & \delta_{4}^{*}
\end{array} \right) \right|_{\rm no\, cross-brane}\label{NeutrinoEff},
\end{equation}
where the entries $\delta_{i}$ are anyway complex due to the $\omega$ factors in
$U(\omega)^{\dagger}$. This matrix is identical to the one
obtained in \cite{Csaki:2008qq}; this is not surprising, since their
model is the limit of our model when $\chi$ is
confined to the UV brane, while $\Phi$ and the Higgs are confined to
the IR brane. One can easily verify that the above matrix is
diagonalized with $\theta_{23}=\pi/4$ and $\theta_{13}=0$,  if
$\delta_2$ and $\delta_4$  are real. We can conclude that for particular realizations of 
the parameters, it is certainly possible that the leading order values  
$\theta_{23}=\pi/4$ and $\theta_{13}=0$ remain protected from higher order brane-localized 
corrections, while in the most general case they will be corrected by naturally suppressed 
contributions.  

The most dominant cross-brane operator is $\overline{\ell}_L\chi H\nu_{R}$. 
If the bulk mass of $\chi$
is vanishing, this operator is suppressed only by
$\epsilon^{\chi}_{13}$, compared to the leading order Dirac mass term.
Keeping the perturbative expansion linear in each of the various
$\epsilon'$s, one can treat the textures associated with each
operator in an additive manner. Therefore, even before we set the
exact profile of $H$ and $\chi$ and deal with all of the other operators, we 
can have a good
understanding of the deviations it induces to $\theta_{12}$,
$\theta_{13}$ and $\theta_{23}$.

We now recall   that in general the  effective Majorana Mass
matrix is a $3\times 3$ complex symmetric matrix and thus contains
$12$ parameters. These parameters are the $3$ masses, the $3$
mixing angles and $6$ phases, out of which $3$ can be absorbed in
the neutrino fields and the remaining are $2$ Majorana and $1$
KM phase. Notice also that, in general, the various
$\epsilon_{ij}^{\chi}$ of Eq.~(\ref{DiracTextures}) are complex.
Thus, considering only the contributions of operators of the
form $\overline{\ell}_L\chi^m H\nu_{R}$, the left-diagonalization
matrix is now corrected to:
\begin{equation}
V_L^{\nu} = \left( \begin{array}{ccc} 1 & 0 & 0 \\ 0 & 1 & 0 \\ 0
& 0 & e^{i\delta}
\end{array} \right)
\left( \begin{array}{ccc} 1/2(\sqrt{2}-\epsilon_{\chi}^2) & \ 0 & -(1/\sqrt{2}+\epsilon_{\chi}) \\
0 & \ 1 & \ 0
\\ 1/\sqrt{2}+\epsilon_{\chi} & \ 0 & \ 1/2(\sqrt{2}-\epsilon_{\chi}^2)
\end{array} \right)
\left( \begin{array}{ccc} e^{i\alpha_1} & 0 & 0 \\ 0 &
e^{i\alpha_2} & 0 \\ 0 & 0 & e^{i\alpha_3}
\end{array} \right),
\label{epschidef}
\end{equation}
where $\epsilon_{\chi} \sim{\cal O}(\chi_0/\Lambda_{5D}^{3/2})$
stands for contributions from $\epsilon_{13}^{\chi}$ and
$\epsilon_{11,22}^{\chi}$ in Eq.~(\ref{DiracTextures}) and we have
omitted terms of $\mathcal{O}(\epsilon_{\chi}^3)$ and higher. 
In particular, $\epsilon_{13}^{\chi}\sim \tilde{y}_\nu^D H_0\epsilon_{\chi}$ and 
 $\epsilon_{11,22}^{\chi}\sim \tilde{y}_\nu^D H_0\epsilon_{\chi}^2$.
The
phases $\alpha_i$ can be absorbed in a rotation of the neutrino fields, while the 
KM phase $\delta$, given by
\begin{equation}
\delta = {\rm Arg}(\tilde{M} + |\tilde{M}|\epsilon^{\chi *}_{11})) - {\rm Arg}(\tilde{M} +
|\tilde{M}|\epsilon^{\chi}_{11}),\label{NuCP}
\end{equation}
will contribute to $CP$ violation in neutrino oscillations.
The MNSP matrix, to $\cal{O}(\epsilon_{\chi})$, acquires a simple structure
\begin{equation}
V_{MNSP} = U(\omega)^{\dagger} V_L^{\nu} = \frac{1}{\sqrt{6}}
\left( \begin{array}{ccc}
(1+e^{i\alpha}\epsilon_{\chi}) & \sqrt{2} & (e^{i\alpha}-\epsilon_{\chi}) \\
(1+\omega e^{i\alpha}\epsilon_{\chi}) & \sqrt{2}\omega^2 & (\omega e^{i\alpha} -\epsilon_{\chi}) \\
(1+\omega^2 e^{i\alpha}\epsilon_{\chi}) & \sqrt{2}\omega &
(\omega^2 e^{i\alpha}-\epsilon_{\chi})
\end{array} \right)\, ,\label{Nufix}
\end{equation}
where the middle column does not receive corrections. A non zero $\theta_{13}$ is generated,
 and
$\theta_{12}$ and $\theta_{23}$ deviate from their leading order
bimaximal values.
Defining $\theta=\pi/4+\epsilon_{\chi}$, the Jarlskog invariant turns out to be
\begin{equation}
{\rm Im}[V_{11}\, V_{12}^*\, V_{21}^*\, V_{22}] =
\frac{\sqrt{3}}{18} \left(\cos 2\theta - \sin 2\theta \sin\delta
\right)\,,
\end{equation}
where the $V_{ij}$ denote the entries of $V_{MNSP}$.

To account for all possible deviations from tribimaximal mixing we  should consider the
fully perturbed effective neutrino mass matrix to first order in
all the $\epsilon$'s defined in Eqs.~(\ref{DiracTextures})
and~(\ref{Majtexture}). However, considering the scales and
wavefunction overlaps associated with $\hat{\epsilon}_{i}$ in
Eq.~(\ref{DiracTextures}) and $\tilde{\epsilon}_{i}$ in
Eq.~(\ref{Majtexture}), it is clear that their contributions are of
characteristic strength $\mathcal{O}(10^{-3})$. Consequently,
the deviations from TBM induced by these effect are negligible and actually below the model 
theoretical error. To
complete our analysis, the only contributions that need to be
further studied are those encoded in $\epsilon_i$ of
Eq.~(\ref{Majtexture}) and those associated with complex
$\epsilon_{i}^{\Phi}$ parameters in Eq.~(\ref{DiracTextures}). The
resulting expressions for the deviations induced by these effects
turn out to be quite cumbersome, yet the largest deviations
 are still almost an order of magnitude smaller than those
described in Eq.~(\ref{Nufix}). These contributions will anyway be
taken into account in the estimations of Sec.~\ref{CrossLep}.

\subsection{Cross-talk and cross-brane effects in the charged fermion
sector}\label{Crosstalk1} 

As suggested in \cite{Volkas}, $Z_3$ breaking cross-talk effects 
should produce deviations of the CKM matrix from unity.
The details of the cross-talk depend on the specific dynamics and
in our case on the wavefunction overlap of $\Phi$ and $\chi$. 
After introducing the higher dimensional cross-talk operators for quarks, 
we will set the values of $\Phi$, $\chi$ and $H$ VEVs to obtain
the physical quark masses while maintaining the Yukawa couplings $y_{u^i, d^i}\simeq 1$.

The higher dimensional 5D  operators relevant for quark mixing are generically
suppressed by $\Lambda_{5D}^{7/2}$. Schematically, they are of the
form
\begin{eqnarray}
& \overline{Q}_L \, u_R \,H\,\Phi \, \chi,\quad \overline{Q}_L \,
u'_R \,H\, \Phi \, \chi,\quad \overline{Q}_L \, u''_R \,H\, \Phi
\, \chi, &
\nonumber\\
& \overline{Q}_L \, d_R \, \tilde{H}\,\Phi \, \chi,\quad
\overline{Q}_L \, d'_R \, \tilde{H}\,\Phi \, \chi,\quad
\overline{Q}_L \, d''_R \, \tilde{H}\,\Phi \, \chi, &
\label{eq:quarkeffops}
\end{eqnarray}
The VEV of the UV peaked $\chi$ communicates $Z_3$ breaking to
the quarks through these operators, and they are suppressed by
an amount $\delta\simeq{\cal O}(\chi_0/\Lambda_{5D}^{3/2})$ compared to the
leading order. They are also
suppressed by the small overlap between the bulk
profiles of $\chi$ and $\phi$, a suppression that amounts to a multiplicative factor, 
slightly different for the various operators.

We focus on the $A_4$ textures associated with the above
operators and conveniently disregard $H$ in what follows, since it transforms
trivially under $A_4$. 
Each operator in Eq.~(\ref{eq:quarkeffops}) represents two independent $A_4$
invariants; for example $\overline{Q}_L \,\Phi \,
\chi\, u_R $ schematically denotes the independent terms
\begin{equation}
[\, (\, \overline{Q}_L\, \Phi\, )_{\3s}\, \chi\, ]_{\s}\, u_R\quad
{\rm and}\quad [\, (\, \overline{Q}_L\, \Phi\, )_{\3a}\, \chi\,
]_{\s}\, u_R. \label{eq:35}
\end{equation}
Writing these terms according to the decomposition in the appendix, and after $A_4$ breaking 
by the VEVs in Eqs.~(\ref{eq:C3vac}) and (\ref{eq:Z2vac}), we obtain the 
general corrections to the 4D quark mass matrices, as also in \cite{Volkas}
\begin{equation}
\Delta M_{u,d} = \left( \begin{array}{ccc}
x_1^{u,d} & \ \ x_2^{u,d} & \ \ x_3^{u,d} \\
0 & \ \ 0 & \ \ 0 \\
y_1^{u,d} & \ \ y_2^{u,d} & \ \ y_3^{u,d}
\end{array} \right)\,,
\label{IDCKM}
\end{equation}
where all  entries are in general complex, and the leading order mass matrix is linearly 
corrected to, $q=u,d$
\begin{eqnarray}
M_q + \Delta M_q & = & U(\omega) \, \sqrt{3}\, \left(
\begin{array}{ccc}
\tilde{y}_{q_1} v + (x_1^q + y_1^q)/3 & \ \ (x_2^q + y_2^q)/3 & \ \ (x_3^q + y_3^q)/3 \\
(x_1^q + \omega y_1^q)/3 & \ \ \tilde{y}_{q_2}' v + (x_2^q + \omega y_2^q)/3 & \ \ (x_3^q + \omega y_3^q)/3 \\
(x_1^q + \omega^2 y_1^q)/3 & \ \ (x_2^q + \omega^2 y_2^q)/3 & \ \
\tilde{y}_{q_3}'' v + (x_3^q + \omega^2 y_3^q)/3
\end{array} \right) \nonumber\\
& \equiv & U(\omega) V_L^{u,d} \left( \begin{array}{ccc} m_{u,d} & 0 & 0
\\ 0 & m_{c,s} & 0 \\ 0 & 0 & m_{t,b}
\end{array} \right) V_R^{u,d \dagger}\,,\label{DCKM}
\end{eqnarray}
with $x_i, y_i\sim{\cal
O}(\tilde{y}_{q_i}v\chi_0/\Lambda_{5D}^{3/2})$.
The left-diagonalization matrices are promoted to $U(\omega) V_L^{u,d}$, 
where $V_L^{u,d}$ are nearly diagonal
for $x_i^q$'s and $y_i^q$'s small enough compared to the mass eigenvalues 
$\sim\tilde{y}v$. Consequently, the CKM matrix will be given by
\begin{equation}
V_{CKM} = \left([\, U(\omega)\, V_L^u\, ]^{\dagger} \, [\,
U(\omega)\, V_L^d\, )]\right) = (V_L^{u\dagger} \, V_L^{d} ) \neq
1.
\end{equation}
There should be  enough
freedom in $V_L^{u,d}$ to fit the observed CKM matrix, while still
explaining why it is nearly the identity. However, it is not obvious that an appealing 
solution can be found, that is an economical 
one, satisfying all perturbativity constraints with a minimal amount of fine tuning.
To account for the more detailed features
of the CKM matrix, like the hierarchy of $\theta_{12}$, $\theta_{23}$ and $\theta_{13}$, 
we will first obtain the  effective 4D couplings
of the above operators, $x_i^{u,d}$ and $y_i^{u,d}$, in terms of
the 5D ones, $\tilde{x}_i^{u,d}$ and $\tilde{y}_i^{u,d}$. Secondly, we search
for the minimal number of parameter assignments
possible, in order to obtain a realistic CKM matrix.

At this point, with all scales fixed in the quark sector, we will need to estimate 
the size of the analogous 
cross-talk effects back in the charged lepton sector, mediated by operators of the form 
$\overline{\ell}_L\Phi H \chi e_R(e_R', e_R'')$, and the way they affect the neutrino 
mixing matrix. 
Clearly, we want to preserve the appealing neutrino mixing pattern described
in the previous section, at least within the $1\sigma$ range of the
experimental values for the neutrino mixing angles~\cite{Fogli:2008}. 
The maximal deviations from TBM will be discussed in Sec.~\ref{CrossLep}.
\section{Numerical results for fermion masses and mixings}
\label{FindCKM} 
We first proceed to set all VEVs and mass scales, according to the observed mass 
spectrum and neutrino oscillation data.  
Once these scales are set, we will be able to quantify the various higher order
contributions, starting from the quark sector.

We take the fundamental 5D scale to be
$k\simeq\Lambda_{5D}\simeq M_{Pl}$, where $M_{Pl}\simeq 2.44\times
10^{18}$ GeV is the reduced Planck mass. To keep the scale of the
IR brane in reach of future collider experiments, but satisfying  the constraints from 
the observed S and T parameters, we will always use values around
$k\pi R\simeq 34$, such that the mass of the first KK excitation
is of a few TeV. It is natural to expect all of the scalars
in our theory, being bulk fields, to acquire a VEV of order $M_{Pl}$. 
Using the known mass of the $W$  boson we can set the
amplitude of  $H$ in terms of the 5D weak gauge couplings to be 
$H_0=0.396M_{Pl}^{3/2}$. 

By exploiting the warped geometry, we can match all the observed 
4D fermion masses by taking the bulk fermion parameters $c_{q,l,\nu}$ to be 
all of order one, a well known pleasing feature of the warped scenario 
\cite{bulkfermion1}; a 
large mass hierarchy is seeded by a tiny hierarchy of the $c$ parameters. 
   
In this setting, fermion masses are determined by the overlap integrals in the 
corresponding Yukawa terms, involving the fermion zero mode profiles  and
the scalar VEV profiles of 
Eqs.~(\ref{profiles}) and (\ref{eqn:ffactor}). Therefore, if we
take the 5D Yukawa couplings to be universal and set them
to one, all bulk parameters can be matched to the observed mass spectrum. 
We remind the reader that there is only one bulk parameter $c_L^{q,l}$ for each 
 left-handed quark and lepton doublet, being it a triplet of $A_4$.
Also, the $c_L^{q,l}$ are essentially
free parameters,  since we can always set the mass of each fermion
by tuning the bulk mass of the corresponding right handed $A_4$
singlet. For the same reasons,
the scalar VEV $\Phi_0$ is essentially also a free parameter and we set it to be $\Phi_0=0.577M_{Pl}^{3/2}$.
However, all fermion zero mode wave functions associated with the
various $c$ parameters are still constrained by a few important
perturbativity bounds~\cite{Agashe:2004cp} and by precision measurements. 
The most stringent constraint is on the quark left-handed bulk parameter $c^q_L$ and comes from the bottom sector, in particular the combined best fits for the ratio of the Z-boson decay width into bottom quarks and the total hadronic width $R_b^0$, the bottom-quark left-right asymmetry, $A_b$, and the forward-backward asymmetry, $A_{FB}^{0,b}$ \cite{ALEPH}.

\subsection{The quark and charged lepton sector}

A thorough comparison of the combined best fits for $R_b^0$, $A_b$, $A_{FB}^{0,b}$ with the  tree-level corrections to the $Zb\bar{b}$ couplings in the minimal (non custodial) RS model can be found in \cite{Casagrande_RSmin}, while the case of $P_{LR}$ custodial symmetry and extended $P_{LR}$ has been recently considered \cite{Casagrande_ECust}.  
These analyses\footnote{  Notice that the analyses of \cite{Casagrande_RSmin, Casagrande_ECust} take into account contributions induced by the complete summation over KK states and their non orthonormality.
These corrections were not considered in previous bounds and turn out to be numerically significant in the case of $Zb\bar{b}$ constraints.} 
show that the allowed window for new physics corrections to the SM prediction -- and consequently the window for the corresponding bottom bulk parameter $c^b_L$ -- is severely constrained by the $Zb\bar{b}$ best fits, unless an extended custodial symmetry such as $P_{LR}$ custodial, or extended $P_{LR}$ custodial is in place.
The model considered here is an intermediate example, which embeds non-extended custodial symmetry, and it is thus expected to be subject to constraints on the left-handed profile $c^q_L$ more severe than in the $P_{LR}$ custodial case and possibly close to the minimal RS model. On the other hand, the model also differs from the cases analyzed in \cite{Casagrande_RSmin,Casagrande_ECust} in two respects. 
The additional $A_4$ flavor symmetry might modify the non orthonormality properties of bulk fermions and eventually suppress contributions due to mixing of zero modes with KK states. Secondly, a bulk Higgs instead of a brane localized Higgs generally allows for further suppressions via overlap in the presence of mass insertions. We defer to future work the analysis of these two aspects in the context of $Zb\bar{b}$.

For the purpose of model building we provide here an approximate estimate of the allowed range for $c^q_L$ in our model, at tree level and in the ZMA, based on the recent calculations in \cite{Casagrande_RSmin,Casagrande_ECust}.
We correct the contributions to the $Zb\bar{b}$ couplings $g^b_{L(R)}$ in the minimal RS model of \cite{Casagrande_RSmin}, 
with the dominant $\omega_Z^{b_{L,R}}\neq 1$ corrections due to the presence of custodial symmetry \cite{Casagrande_ECust}. Analogously to \cite{Casagrande_RSmin}, we conveniently define the functions $f(c)$ in terms of the canonically normalized fermion zero mode wave functions $f^{(0)}(y)$ of eq.~(\ref{eqn:ffactor}) evaluated at the IR brane
\begin{equation}
 f^{(0)}(y=\pi R) = \sqrt{k\pi R}\, e^{\frac{3}{2}k\pi R}\,f(c)\, .
\end{equation}
We obtain 
\begin{eqnarray}
g^b_L&=&  \left ( -\frac{1}{2}+\frac{s_w^2}{3}\right ) \left [ 1-\frac{m_Z^2}{2M_{KK}^2}\frac{f^2(c^q_L)}{3-2c^q_L}
\left (\omega_Z^{b_L}\cdot k\pi R - \frac{5-2c^q_L}{2(3-2c^q_L)}\right ) \right ]       \nonumber\\
&&+\frac{m_b^2}{2M_{KK}^2}\left [\frac{1}{1+2c_b} \left (\frac{1}{f^2(c_b)}-1+\frac{f^2(c_b)}{3-2c_b}\right )  
+\sum_{i=d,s} \frac{|(Y_d)_{3i}|^2}{|(Y_d)_{33}|^2} \frac{1}{1+2c_i}\frac{1}{f^2(c_b)}   \right ] \nonumber\\
g^b_R&=&  \frac{s_w^2}{3}\left [ 1-\frac{m_Z^2}{2M_{KK}^2}\frac{f^2(c_b)}{3-2c_b}
\left (\omega_Z^{b_R}\cdot k\pi R - \frac{5-2c_b}{2(3-2c_b)}\right ) \right ]       \nonumber\\
&&-\frac{m_b^2}{2M_{KK}^2}\left [\frac{1}{1+2c^q_L} \left (\frac{1}{f^2(c^q_L)}-1+\frac{f^2(c^q_L)}{3-2c^q_L}\right )  
+\sum_{i=d,s} \frac{|(Y_d)_{3i}|^2}{|(Y_d)_{33}|^2} \frac{1}{1+2c^q_L}\frac{1}{f^2(c^q_L)}   \right ]   \, 
\label{eq:Zbb}
\end{eqnarray}
where $\omega_Z^{q} = c_w^2 (T_L^{3q}+T_R^{3q})/(T_L^{3q}-s_w^2 Q_q)$ in the case of equal $SU(2)_L$ and $SU(2)_R$ gauge couplings.
It is instructive to compare the values for $\omega_Z^{b_{L,R}}$ in the three different setups. The minimal RS model has $\omega_Z^{b_{L}}=\omega_Z^{b_{R}}=1$, the $P_{LR}$ custodial has  $\omega_Z^{b_{L}}=0$ and $\omega_Z^{b_{R}}=3c_w^2/s_w^2\sim 10$ for $s_w^2\approx 0.23$, while the custodial case as in our model has $\omega_Z^{b_{L}}= c_w^2/(1-2s_w^2/3)\sim 0.9$ and $\omega_Z^{b_{R}}=-3c_w^2/2s_w^2\sim -5$. Notice that the latter is negative, and would go in the right direction to solve the $A_{FB}^{0,b}$ anomaly. We have however verified that its numerical impact is limited, as it is its positive contribution in the minimal RS setup. 

In the estimate provided by eq.~(\ref{eq:Zbb}) we have disregarded corrections to the $m_b$ dependent terms due to the admixture of the KK partners in the zero modes of the $SU(2)_R$ doublets; the exact form of the corrections depends on the symmetries of the model, also $A_4$ flavor in our case, and weak isospin assignments of bulk fermions. The $m_b$ dependent contribution as in eq.~(\ref{eq:Zbb}) 
is strongly dependent on the right-handed bottom bulk paramater $c_b$. However, it can be effectively suppressed 
in the particular case of extended $P_{LR}$ custodial symmetry due to degeneracy of right-handed profiles \cite{Casagrande_ECust}.

 We used the combined best fit values for the couplings $(g_L^b)_{exp}=-0.41918$ and $(g_R^b)_{exp}=0.090677$ \cite{Casagrande_RSmin} and the predicted SM values 
$(g_L^b)_{SM}=-0.42114$ and $(g_R^b)_{SM}=0.077345$ \cite{ALEPH} at the reference Higgs mass of 150 GeV, to constrain the new physics contributions defined as $\delta g^b_{L(R)} = g^b_{L(R)} -(g^b_{L(R)})_{SM}$.
We disregarded the new physics corrections to the SM contributions in the light sector as in \cite{Casagrande_RSmin, Casagrande_ECust}. 

\begin{figure}
\begin{center}
\includegraphics[width=8 truecm]{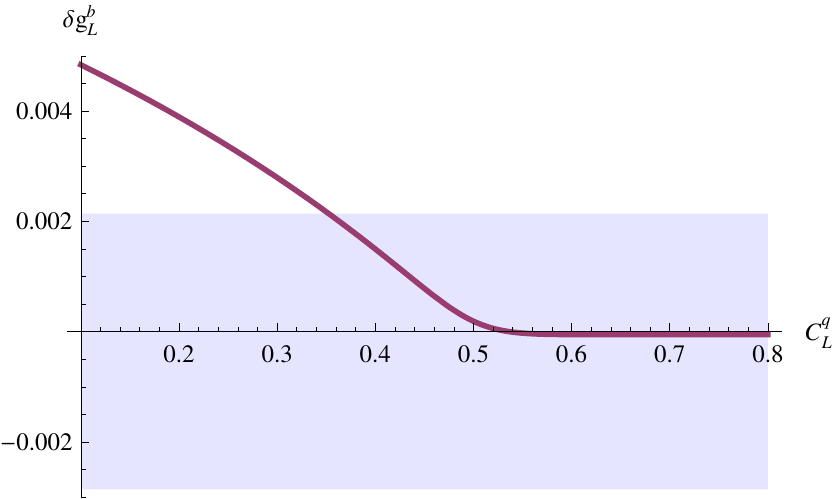}
\caption{\label{fig:zbbconstraint} An illustrative example of the new physics contribution $\delta g_L^b$ as a function of $c^q_L$, with $M_{KK}=1.8$ TeV, the down-type 5D Yukawa couplings in eq.~(\ref{eq:Zbb}) set to 1 in magnitude and the right-handed bulk parameters set to the conservative value $c_b=0.58$, and $c_d=c_s=0.5$. The shaded horizontal band corresponds to the $99\%$ probability interval allowed by the best fit values for $g^b_{L(R)}$ with the SM prediction computed at $m_H=150$ GeV as in \cite{Casagrande_RSmin, Casagrande_ECust}.  }
\end{center}
\end{figure}

For a Higgs mass of 150 GeV, a KK scale $M_{KK}=1.8$ TeV, and imposing a $99\%$ probability interval for the left-handed coupling 
given by $-0.424\lesssim g^b_{L}\lesssim -0.419$, we obtain the constraint $c^q_L>0.35$, as it can be inferred from figure \ref{fig:zbbconstraint}. Notice that eq.~(\ref{eq:Zbb}) has a strong dependence on $c_b$ and only a mild dependence on $c_{d,s}$. Thus, lowering the value of $c_b$ towards 0.5 will allow for significantly larger windows for $c^q_L$, at the price of higher Yukawa couplings. The corrections to the SM prediction for $g_R^b$ satisfy $\delta g^b_R\lesssim 10^{-4}$ in the region of interest and have been safely neglected.
The perturbativity constraint on the 5D top Yukawa coupling $|y_t|\lesssim\pi$, together with the matching to the 
$\overline{\mbox{MS}}$ top mass at the KK scale $m_t(1.8\,\mbox{TeV})\simeq 140\,\mbox{GeV}$ \cite{PDG} and the constraint
$c_t\gtrsim -0.2$\footnote{This constraint ensures that the contribution through mixing from $\tilde{b}$, the partner of the $SU(2)_R$ top, is sufficiently suppressed in $Zb\bar{b}$ \cite{Agashe:2004cp,Agashe:2003zs}.}, implies instead the upper bound $c^q_L<0.42$. Notice that by requiring $|y_t|\lesssim 4\pi /\sqrt{N}$ \cite{Burdman:2003ya}, with $N=1$, the bound on $c^q_L$ changes significantly to $c^q_L<0.52$. Hence, one obtains $0.35\lesssim c^q_L\lesssim 0.42$ in the most conservative case.
As expected, the allowed minimal value for $c^q_L$ is severely constrained by the $Zb\bar{b}$ best fits.

For $c^q_L=0.40$, all the right-handed quark bulk parameters are chosen in order to yield the $\overline{\mbox{MS}}$ quark masses at 
the KK scale of 1.8 TeV.
Their values are $c_u=0.790$, $c_d=0.770$, $c_s=0.683$, $c_c=0.606$, $c_b=0.557$ and $c_t=-0.17$, 
corresponding to $m_u=1.5\,\mbox{MeV}$, $m_d =3\,\mbox{MeV}$, $m_s=50\,\mbox{MeV}$, $m_c=550\,\mbox{MeV}$, $m_b=2.2\,\mbox{GeV}$ and $m_t=140\,\mbox{GeV}$. All 5D Yukawa couplings have been set to 1 in magnitude, while the top Yukawa coupling slightly breaks universality with $|y_t|\simeq 2.8$.

It should be added that lower values for $c_L^q$ become accessible for a KK scale higher than $M_{KK}=1.8$ TeV. On the other hand, 
a KK scale as low as 1 TeV would force $c^q_L\gtrsim 0.45$ and the top Yukawa coupling to values $|y_t|\sim 7$. Another possibility is the one of a heavier Higgs.
A Higgs mass larger than 150 GeV would bring the model prediction closer to the best fit values for $g_L^b$ and $g_R^b$, thus allowing for a larger range for $c^q_L$. 
For example, a mass $m_H =300$ GeV would imply a lower bound $c_L\gtrsim 0.32$, within our estimate and using 
the shifts $\Delta g_L^b = 1.77\cdot 10^{-3}\ln\left ({m_H}/{150\mathrm{GeV}}\right )$ and $\Delta g_R^b = 0.92\cdot 10^{-2}\ln\left ({m_H}/{150\mathrm{GeV}}\right )$ \cite{Casagrande_ECust} induced by $m_H\neq 150$ GeV. However, a heavier Higgs mass in the custodial setup easily induces conflicts with electroweak precision measurements and a careful estimate of the actually allowed range of values for $m_H$ should be produced in each version of the custodial setup. We defer this point to future work on phenomenological applications.
 
For the charged leptons we make the  choice $c_L^l=0.52$,
$c_e=0.803$, $c_\mu=0.635$ and $c_\tau=0.5336$, which reproduce the experimental value of the corresponding masses
$m_e=0.511$ MeV, $m_\mu=105.6$ MeV and $m_\tau=1.776$ GeV.

\subsection{The neutrino sector}
%%%
To obtain the neutrino masses we first recall the leading order
structure of the light
neutrino mass matrix after see-saw, $M^\nu_L$ in Eq.~(\ref{eq:MLbare}), for
which the eigenvalues  are given by:\begin{equation}
M_L^{diag.}=-(m_\nu^D)^2\times\left[\frac{1}{\tilde{M}+M_\chi},
\frac{1}{\tilde{M}},
\frac{1}{\tilde{M}-M_\chi}\right]\, .
\label{eigen}
\end{equation} 
We now write the neutrino mass-squared splittings, for afterwards we
would like to impose the observed values of $\Delta m_{{\rm
atm}}^{2}$ and $\Delta m_{{\rm sol}}^{2}$, in order to constrain
the possible choices of $\tilde{M}$ and $M_{\chi}$. Given 
\begin{equation}
\Delta m_{12}^{2}\equiv |m_1|^2-|m_2|^2=
\left|\frac{(m_{\nu}^{D})^{2}}{(\tilde{M})}\right|^{2}\left[\frac{1}
{(1+q)^{2}}-1\right]\label{ratio1}\,,
\end{equation}
\begin{equation}
\Delta m_{23}^{2}\equiv |m_2|^2-|m_3|^2=
\left|\frac{(m_{\nu}^{D})^{2}}{(\tilde{M})}\right|^{2}\left[1-\frac{1}{(1-q)^{2}}\right]
\label{ratio2},
\end{equation}
where $q=M_{\chi}/\tilde{M}$, $|\Delta m_{12}^2|=\Delta
m_{sol}^{2}$ and $|\Delta m_{23}^2|=\Delta m_{atm}^{2}$, we
 obtain the following cubic equation
for $q$:
\begin{equation}
q^{3}-3q-2\left(\frac{x-1}{x+1}\right)=0\, ,
\label{cubic}
\end{equation}
where $x=\Delta m_{sol}^{2}/\Delta m_{atm}^{2}$  for $|q|<2$, and
$x\rightarrow -x$ for $|q|>2$. Given $q$, the ratio $\tilde{M}/(m_{\nu}^{D})^{2}$
can be extracted either in terms of $\Delta m_{atm}^{2}$ or $\Delta m_{sol}^{2}$
\begin{equation}
\tilde{M}=\frac{(m_{\nu}^{D})^{2}}{\sqrt{\Delta
m_{sol}^{2}}}\times\left(\left|\frac{1}{(1+q)^{2}}-1\right|\right)^{-1/2}.
\end{equation}
Imposing  the measured values \cite{Fogli:2008} for the mass
splittings, $\Delta m^2_{sol}\simeq 7.67\times10^{-5}{\rm eV}^{2}$
and $\Delta m^2_{atm}\simeq 2.39\times 10^{-3}{\rm eV}^{2}$, we
find four possible solutions to the above equation,
$q\simeq\left\{-2.02, -1.99, 0.79, 1.2\right\}$, where the first two
correspond to inverted hierarchy, while the second two correspond
to normal hierarchy. In addition, once we set $c_{\nu_R}$, we can
constrain $\tilde{M}$ and $M_{\chi}$ from the same data. This is a
nice feature of the light neutrino mass matrix obtained in all
$A_4$ models with similar
assignments~\cite{a4,Csaki:2008qq,Volkas}.
Since at this stage only the overall light
neutrino mass ratios, $(m_{\nu}^{D})^{2}/\tilde{M}$ and $M_\chi /\tilde{M}$ are
constrained by the observed splittings, we choose not to set
$c_{\nu_R}$ and extract $M_\chi$ from the quark mixing
data. It will be possible afterwards to set
$c_{\nu_R}$ to a natural value $\sim 1/2$ to match the neutrino mass spectrum.
%%%%%%%%%%%%%%

\subsection{Obtaining the CKM matrix and fixing the scale} 
\label{MatchCKM} 
We now analyze the CKM matrix resulting
from the contributions introduced in Eq.~(\ref{DCKM}). Using
standard perturbative techniques, the left diagonalization matrix 
$V_{L}^{u,d}$ is obtained by the unitary diagonalization
of $(M+\Delta M)_{u,d}(M+\Delta M)^{\dagger}_{u,d}$,
and the right diagonalization matrix $V_R^{u,d}$ is analogously obtained by 
the unitary diagonalization of $(M+\Delta M)_{u,d}^{\dagger}(M+\Delta M)_{u,d}$.
The entries of the CKM matrix are then
derived in terms of the $x_{i}^{u,d}$, $y_{i}^{u,d}$ parameters
defined in Eq.~(\ref{DCKM}), to leading order in perturbation
theory. The up and down left-diagonalization matrices turn out to
be:
\begin{equation}V_L^q=\left( \begin{array}{ccc} 1 &
m_2^{-1}(x_2^q+y_2^q) &
m_3^{-1}(x_3^q+y_3^q)\\
-m_2^{-1}(\bar{x}_2^q + \bar{y}_2^q) & 1 &
m_3^{-1}(x_3^q + \omega y_3^q)\\
-m_3^{-1}(\bar{x}_3^q + \bar{y}_3^q)& -m_3^{-1}(\bar{x}_3^q +
\omega^2\bar{y}_3^q)& 1
\end{array}\right)\label{VLQ},
\end{equation}
with $q=u,d$, $m_i=m_{u_i, d_i}$, and $\bar{x}(\bar{y})$ stands for the complex conjugate. 
In the above matrices, for simplicity, we have redefined the
$x_i^{u,d}$ and $y_i^{u,d}$ to absorb the relative factor of
$1/\sqrt{3}$ compared to the unperturbed mass eigenvalues
$m_i=\sqrt{3}v\tilde{y}_{u^i, d^i}$. Furthermore, we have omitted
contributions that are suppressed by
quadratic quark mass ratios and still linear in $x_i^{u,d}$,
$y_i^{u,d}$.
Terms of this kind will be included in the complete expressions derived below, for each of
 the interesting CKM matrix elements. 

It is now straightforward to extract the estimations for the upper
off-diagonal elements of the CKM matrix out of
$V_L^{u\dagger}V_L^d$, in order to eventually match the three
mixing angles and the CP violating phase $\delta^{13}_{CKM}$. To
leading order in $x_i,\, y_i$, these elements turn out to be
\begin{equation}
V_{us} \simeq -V_{cd}^{*}\simeq \frac{m_u(\bar{x}_1^u + \omega^2
\bar{y}_1^u)+ m_c(x_{2}^{u}+
y_{2}^{u})}{m_u^2-m_c^2}+\frac{m_d(\bar{x}_{1}^{d} + \omega^2
\bar{y}_{1}^{d})+m_s(x_2^d+y_2^d)}{m_s^2-m_d^2} \label{shoot1}\,,
\end{equation}
\begin{equation}
V_{cb}\simeq -V_{ts}^{*}\simeq \frac{m_c(\bar{x}_2^u+\omega
\bar{y}_2^u)+ m_t(x_{3}^{u}+ \omega
y_{3}^{u})}{m_c^2-m_t^2}+\frac{m_s(\bar{x}_{2}^{d}+\omega
\bar{y}_{2}^{d})+m_b(x_3^d+\omega y_3^d)}{m_b^2-m_s^2}
\label{shoot2}\,,
\end{equation}
\begin{equation}
V_{td}\simeq -V_{ub}^{*}\simeq \frac{m_u(\bar{x}_1^u+\omega
\bar{y}_1^u)+ m_t(x_{3}^{u}+
y_{3}^{u})}{m_u^2-m_t^2}+\frac{m_d(\bar{x}_{1}^{d}+\omega
\bar{y}_{1}^{d})+m_b(x_3^d+ y_3^d)}{m_b^2-m_s^2}\label{shoot3}\,,
\end{equation}
where it is important to observe that the first equality is exact
to leading order in $x_i$ and $y_i$. The diagonal elements of the
CKM matrix remain unchanged at this order and equal to one.

We recall that $x_i^{u,d}$ and $y_i^{u,d}$  correspond to the
$(1i)$ and $(3i)$ entries of $\Delta M^{u,d}$ in the interaction
basis, respectively. This tells us which fermionic wave function
overlaps enter the integral for each of the above parameters.
For the 4D couplings we thus obtain
\begin{equation}
x_{i}^{u,d}(y_i^{u,d})=\left(\frac{H_0\Phi_0\chi_0}{\Lambda_{5D}^{7/2}}\int_{-\pi
R}^{\pi R}\frac{dy}{2\pi R}\, F(c_L^q, c_i^{u,d})e^{8(k|y|-\pi R)}(1-e^{4(k|y|-\pi
R)})\right)\tilde{x}_{i}^{u,d}(\tilde{y}_i^{u,d})\, .\label{scalingX}
\end{equation}
To narrow down the parameter space, we choose all the
$\tilde{x}_{i}^{u,d}(\tilde{y}_i^{u,d})$ to be universal and equal
to one in magnitude, while relative phases between these parameters
will be allowed. Hence, due to the
mass hierarchy of the quarks, which keeps the denominators in
Eqs.~(\ref{shoot1})\,-\,(\ref{shoot3}) proportional to just one
quark mass to a good approximation, the resulting corrections to each of the CKM matrix
elements are of the generic form,
\begin{equation}
\frac{m_i^{u,d}(x_{i}^{u,d}+\omega^{n}y_{i}^{u,d})}{(m_j^{u,d})^2-(m_k^{u,d})^2}\Rightarrow
\frac{((m_i^{u,d})^2C_{\chi}f_{\chi}^{\,i})(\tilde{x}_{i}^{u,d}+\omega^{n}\tilde{y}_i^{u,d})}{\pm
max[(m_j^{u,d})^2,
(m_k^{u,d})^2]}\label{GenericCKM}\,,
\end{equation}
with $n=0,1,2$, and  where
$C_{\chi}=\chi_0/M_{Pl}^{3/2}$,
$f_{\chi}^i=4/(12-c_L^q-c_i^{u,d})$ and $i=j$ or $k$. Let us first
set $C_{\chi}$ according to the experimental value of $V_{us}$;
using  Eq.~(\ref{shoot1}) and taking into account Eq.~(\ref{GenericCKM}) we
obtain:
\begin{equation}
V_{us}\simeq\left((\tilde{x}_2^d+\tilde{y}_2^d)f_{\chi}^s-(\tilde{x}_2^u+\tilde{y}_2^u)
f_{\chi}^c
+{\cal O}(m_d^2/m_s^2)\right)C_\chi\simeq 0.2257\Rightarrow
C_{\chi}\simeq 0.155 .\label{cchi}
\end{equation}
in which we have fixed $\tilde{x}_2^d$, $\tilde{y}_2^d$,
$\tilde{x}_2^u$ and $\tilde{y}_2^u$ to be $1$ in magnitude,
 with a relative phase, $\delta_2^{u}=\pi$, between the
contributions from the up and down sectors.
\begin{table}[t]
\begin{displaymath}
\begin{array}{|c|c|c|c|c|c|} \hline
      q      & m_1   & m_2  & m_3  & \tilde{M}/M_{Pl} & C_{\chi} \\ \hline
-2.02 &  50.7     &  51.8    &  17.1 & -0.077 & 0.155\\ \hline
-1.99 & 52.3       & 51.8    &   17.3  & -0.077 & 0.155 \\ \hline
 0.79 &  5.8      & 10.5   &    50   & 0.202  & 0.155\\ \hline
 1.19 &  4.3     & 9.4    &    49.4   &  0.135 & 0.155\\ \hline
\end{array}
\end{displaymath}
\caption{Approximate numerical values of the neutrino masses and
relevant UV scales, where $c_{\nu_R}=0.408$ has been chosen to match
the four solutions for $q\equiv M_{\chi}/\tilde{M}$ to Eq. \ref{cubic} with the
constraint $C_{\chi}=0.155$ arising from $V_{us}$. The masses are
given in units of $10^{-3}$ eV.} \label{tab:egmasses}
\end{table}
We want to see if the above value of $C_{\chi}$ together with a
minimal number of relative phases between the remaining
$\tilde{x}$ and $\tilde{y}$ parameters, are enough to account for
the magnitudes of the remaining observed CKM elements. Yet, we first want to
check the consistency of the scale associated with the above value
of $C_{\chi}$ with the neutrino mass splittings and the bare
Majorana mass scale. It turns out that for $c_{\nu_{R}}=0.408$ it
is possible to satisfy the constraints in both sectors, namely to
have a realistic $V_{us}$, while at the same time having a
realistic neutrino mass spectrum with a normal or inverted
hierarchy. At this level the neutrino mixing matrix is obviously
tribimaximal with small deviations, as we saw in section
\ref{secHO}. The numerical results are reported in Table
\ref{tab:egmasses}, once the scales are fixed by $C_\chi$.
The dominant contributions to $V_{cb}$ (and $V_{ts}$) are given by
\begin{equation}
V_{cb}\simeq\left((\tilde{x}_3^d+\omega\tilde{y}_3^d)f_{\chi}^b-(\tilde{x}_3^u+
\omega\tilde{y}_3^u)f_{\chi}^t+{\cal
O}(m_s^2/m_b^2)\right)C_\chi\simeq 0.004\Leftarrow (\delta_3^u=0)\, .
\label{Vcb}
\end{equation}
Up till now, we introduced only one phase, $\delta_2^u$, to match
$V_{us}$ (and $V_{cd}$), but  failed to match $V_{cb}$ (and
$|V_{ts}|$) to their central experimental values
$|V_{ts}|\simeq|V_{cb}|=0.0415$. 
The leading order contribution to $V_{ub}$, without any relative
phase assignments, and with the $\tilde{x}_3^{u,d}$'s and
$\tilde{y}_3^{u,d}$'s set to 1, is
\begin{equation}
V_{ub}\simeq\left((\tilde{x}_3^d+\tilde{y}_3^d)f_{\chi}^b-(\tilde{x}_3^u+
\tilde{y}_3^u)f_{\chi}^t+{\cal O}(m_d^2/m_b^2)\right)C_\chi\simeq
0.007\Leftarrow (\delta_3^u=0)\, .
\label{Vub}
\end{equation}
We see that $V_{ub}$, and thus $V_{td}$, turn out correctly to be of order
$\lambda_{CKM}^3$, but still outside the experimental error.

The next to leading order corrections to the various CKM elements
enter at ${\cal O}((x_i^{u,d}, y_i^{u,d})^{2})$. In general, one
may still expect them to  modify the relatively small values of
$V_{ub}$ and $V_{cb}$, especially in the presence of strong
cancellations at leading order, and given that  each independent
contribution is effectively suppressed  by $f_{\chi}^i
C_{\chi}\approx 0.05$ compared to the corrections linear in
$x_i^{u,d}$ and $y_i^{u,d}$. 
We are going to elaborate on this possibility
once we made an attempt to obtain an almost realistic CKM
matrix, using the first order results.

The only way to obtain a realistic prediction for the independent
magnitudes of $V_{ub}$ and $V_{cb}$, and  the related values of
$|V_{td}|$ and $|V_{ts}|$,  is to break the universality
assumption for the Yukawa couplings, $\tilde{x}_{3}^{u,d}$,
$\tilde{y}_{3}^{u,d}$ and find their values that match the
experimental data. However, we are interested in the smallest
possible deviations from the universality assumption which assumes
all of the Yukawa couplings to be of order one, so that
contributions to various flavor violating processes, which are
present in any RS-flavor setup \cite{rsgim1}, will not be
arbitrarily modified. Therefore, while trying to find the minimum
number of assignments in the $\tilde{x}_{3}^{u,d}$,
$\tilde{y}_{3}^{u,d}$ parameter space, we still require small deviations, in general 
complex, from the $\mathcal{O}(1)$ universality assumption. It is
obvious that assignments in terms of one parameter will yield
results proportional to those of Eq.~(\ref{Vcb}) and (\ref{Vub}),
and will hence fail again to account for realistic values of
$V_{cb}$ and $V_{ub}$. The minimal viable choice consists of at
least 2 parameter assignments. Namely, we will have to break the
universality assumption for two out of the  four
$\tilde{x}_{3}^{u,d}$, $\tilde{y}_{3}^{u,d}$ parameters.

In addition the only choice of parameter assignments that
maintains all  coefficients of $\mathcal{O}(1)$ is to break
universality for $\tilde{x}_3^u$ and $\tilde{y}_3^d$, while
setting $\tilde{x}_3^d=\tilde{y}_3^u=1$. Solving Eqs.~(\ref{Vub})
and (\ref{Vcb}) for $\tilde{x}_3^u$ and $\tilde{y}_3^d$ we
obtain:
\begin{equation} 
\tilde{x}_3^u\simeq 0.67-0.19i,\qquad
\tilde{y}_3^d\simeq 0.60-0.23i \, .
\label{Fassignment}
\end{equation}
These parameters are almost degenerate, in particular if one
considers the overall accuracy of the zero mode approximation, and substituting these 
values in Eqs.~(\ref{Vcb}) and (\ref{Vub}), we obtain
\begin{equation} 
|V_{cb}|=0.0415 \qquad |V_{ub}|=0.0039\, ,
\end{equation} 
which are the central experimental values
~\cite{PDG}. Interestingly, we are able to match also the CP
violating phase $\delta_{13}$, using the same assignment. We find
it to be $\delta_{13}\simeq{1.2}$, which is well within the
experimental error\footnote{At a higher level of accuracy one should obviously satisfy the full set of constraints implied by the measured Jarlskog invariant.}. Obviously, at this order $|V_{ts}|=|V_{cb}|$ holds exactly, and this 
gives
a value for  $|V_{ts}|$ close to the central experimental value
$|V_{ts}|=0.0407$.  We also obtain $|V_{td}|=|V_{ub}|$, and thus fail to match the
central experimental value $|V_{td}|=0.0087$.
Notice, however, that corrections of the size of the smallest CKM entries, i.e. 
$O(\lambda_{CKM}^3)$ are below the model theoretical error induced by the zero mode 
approximation.
A realistic value for $V_{td}$ can easily arise from the subleading corrections in 
$\tilde{x}_{i}^{u,d}$ and $\tilde{y}_i^{u,d}$ and from higher dimensional operators 
beyond the zero mode approximation. We remind that the  corrections quadratic in 
$\tilde{x}_{i}^{u,d}$ and $\tilde{y}_i^{u,d}$ are generically suppressed by a factor 
$f_{\chi}^i C_{\chi}\approx 0.05$ with respect to the linear contributions.

We have thus shown that at leading order in the VEV expansion the model predicts $V_{CKM}$ 
to be the unit matrix, a rather good first step in the description of quark mixing. At the 
next to leading order, cross-brane and cross-talk operators induce deviations from unity, 
parametrized in terms of twelve complex parameters $\tilde{x}_{i}^{u,d}$ and 
$\tilde{y}_i^{u,d}$, with $i=1,2,3$.
We have shown that, linearly in these parameters, realistic values of the CKM entries can 
be obtained within the model theoretical error in a finite portion of the parameter space, 
with all the $\tilde{x}_{i}$ and 
$\tilde{y}_i$ of order one and two non zero relative phases; however at this order, no 
corrections to the diagonal unit entries is produced and the two smallest entries 
$V_{ub}$ and $V_{td}$ are degenerate. 

We notice that the possibility of producing hierarchical CKM entries, of order 
$\lambda_{CKM}$, $\lambda_{CKM}^2$ and $\lambda_{CKM}^3$, with all parameters of order 
one stems from the 
presence of built-in cancellations induced by the hierarchical masses. 
The presence of the $A_4$ induced phase $\omega$ also produces a pattern in the corrections.
In this framework, subleading corrections of order 
$(\tilde{x}_{i}^{u,d},\tilde{y}_i^{u,d})^2$ and contributions beyond the ZMA, 
must be responsible for the deviation from one of the CKM diagonal entries and the non 
degeneracy of $V_{ub}$ and $V_{td}$. The first is of order $10^{-2}$, the latter of order 
$10^{-3}$, hence a cancellation pattern must again be in place. 

It is instructive to compare this $A_4$ pattern with other flavor symmetry groups, in 
particular $T^\prime$ \cite{TPrime}. The Wolfenstein parametrization would 
suggest the 
existence of an expansion parameter to be naturally identified with $\lambda_{CKM}$, 
offering an elegant and simple description of quark mixing in the standard model.
On the other hand, flavor models based on otherwise appealing discrete flavor symmetries 
such as $A_4$ and $T^\prime$ must rely on more complicated patterns to produce a realistic 
CKM matrix. In the case of $T^\prime$ one needs to postulate a hierarchy of Yukawa couplings 
to distinguish between $O(\lambda_{CKM} )$ and $O(\lambda_{CKM}^2 )$ entries, and 
a hierarchy of specific VEVs to induce the splitting between $V_{ub}$ and $V_{td}$.
Similar hierarchies can in principle also be postulated in the $A_4$ case, at the price of 
an increased fine tuning of the input parameters.

\subsection{Estimation of cross-talk and cross-brane contributions in the  lepton
sector}
\label{CrossLep}
To complete our analysis we want to ensure that the deviations
from TBM  induced by cross-talk operators of the
characteristic forms $\overline{\ell}_L\Phi H\chi e_R(e_R',
e_R'')$ and $\overline{\ell}_L H\chi \nu_{R}$  are suppressed
and keep the predictions for the  neutrino mixing parameters
within the $2\sigma$ range of the experimental error. We first
consider the cross-brane operator ${\bar \ell}_L H\chi\nu_{R}$, already
explored in section III. Recall that deviations from
$\theta_{13}=0$ and $\theta_{23}=\pi/4$ are induced only for
complex $\epsilon_{ij}^{\chi}$ parameters, and therefore we expect
the maximal deviations from TBM when they are purely imaginary. All
said, we estimate the magnitude of the
$\epsilon_{ij}^{\chi}$ coefficients based on the results of the
previous section. As we are interested only in the dominant
contributions, associated with one insertion of $\chi$, we are
only interested in $\epsilon_{13}^{\chi}$ as defined in Eq.~(\ref{DiracTextures}). 
This provides 
\begin{equation} \epsilon_{13}^{\chi}\simeq
\tilde{y}_{\nu}H_0
C_{\chi}\frac{4}{8-c^{e}_{L}-c_{\nu_R}}\Rightarrow
\epsilon_{\chi}\simeq 0.07 \, , \label{epsilonChi}   
 \end{equation}
where $\epsilon_{\chi}$ was introduced in Eq.~(\ref{epschidef}). 
Given $\epsilon_\chi$ we also 
obtain $ \epsilon_{11}^{\chi}$. Taking
its phase  to be $\pi/2$ we get the largest possible contribution to the phase 
$\delta$ defined in Eq.~(\ref{NuCP}) to be $\delta\simeq -0.11$.
Using Eq.~(\ref{Nufix}) we can now estimate the deviations from
TBM arising from the operator
$(1/\Lambda_{5D}^{2})\overline{\ell}_L H\chi\nu_R$ in the worst scenario
 where its coefficient is purely imaginary. These
turn out to be 
\begin{equation} \Delta\theta_{13}\simeq 0.05
\qquad \Delta\theta_{23}\simeq 0.04 \qquad \Delta\theta_{12}\simeq
0\,  .
\label{nucrosstalk}
\end{equation} 
All of these deviations are within $1\sigma$ from their experimental values.
The other class of operators $(1/\Lambda_{5D}^{7/2})\overline{\ell}_L\Phi H\chi
e_R(e_R^{'},e_R^{''})$ have the same structure as the
operators in charge of quark mixing. The perturbative
diagonalization procedure can thus be carried out as it is done in the quark
sector, in order to determine the deviations from $U(\omega)$ of the 
rotation matrix for the left handed charged leptons.
Generically, these operators will induce
perturbations to the mass matrix in the form of Eq.~(\ref{DCKM})
and with characteristic strength $\epsilon_{\ell}\simeq C_{\chi}
f_{\chi}(c_{L}^{e}, c_{\ell})\simeq 0.028$, which is of the same
order of the model theoretical error. We could have proceeded to
explicitly write all of  these corrections as we did for the
quarks, yet since the structure is practically identical in both
cases we can easily deduce the effect of these small terms. For
example, in analogy with Eqs.~(\ref{cchi}) and (\ref{Vub}) and using
no additional phase assignments for $\tilde{x}_{i}^e$ and
$\tilde{y}_i^e$, we get that the contribution to
$\Delta\theta_{13}$ is vanishing, while
the contributions to $\Delta\theta_{12}$ and $\Delta\theta_{23}$ are of approximate
strength $|\Delta\theta_{12}|\simeq|\Delta\theta_{23}|\simeq 0.04$, comparable to 
the model theoretical error. Higher order cross-talk
operators are not considered, as their contribution lies safely
below $1\%$. We can conclude that the most significant
deviations from TBM induced by cross-talk and cross-brane operators on the three 
mixing angles stay
within the experimental errors for these quantities. This can be obtained without making any further 
assumption on the parameters of the model and maintaining all of them naturally of order one.

\section{Vacuum Alignment}
\label{Alignment} 
We are still left with one important challenge: to
ensure that the alignments of the VEVs
$\langle\Phi\rangle$ and $\langle\chi\rangle$ in
Eqs.~(\ref{eq:C3vac}) and (\ref{eq:Z2vac}) are
approximately preserved also in the presence of higher order corrections. These VEV 
patterns were the key ingredient
 in obtaining all of the above results. In this section, we
briefly discuss the challenge and suggest possible solutions.  It
is obvious that in the scenario of \cite{Csaki:2008qq}, which
is the limit of our model when $\chi$ and $\Phi$ are strictly localized on the UV and IR brane respectively,
the vacuum alignment problem is eliminated {\em tout court}, as $\Phi$ and $\chi$
are completely sequestered. In our model, the cross-brane interactions between
$\Phi$ and $\chi$ are essential in order to obtain realistic results in the
quark sector. The question is whether their modifications to the scalar potential can
be sufficiently suppressed, while still protecting the results in the lepton and quark 
sectors.

The complete $\mathop{\rm G}$-invariant scalar potential
in $\Phi$ and $\chi$ up to quartic order is displayed below, and we conveniently separate 
its terms as follows 
\begin{equation}
V = V(\Phi) + V(\chi)  + V(\Phi,\chi),
\end{equation}
with the single contributions also derived in \cite{Volkas}
\begin{eqnarray} V(\Phi) &=& \mu^2_\Phi
(\Phi \Phi)_{\s} +\lambda^\Phi_1 (\Phi \Phi)_{\s}(\Phi \Phi)_{\s}
+ \lambda^\Phi_2
(\Phi \Phi)_{\s'}(\Phi \Phi)_{\s''}\nonumber\\
&+&\lambda^\Phi_3 (\Phi \Phi)_{\3s}(\Phi \Phi)_{\3s} +
\lambda^\Phi_4 (\Phi
\Phi)_{\3a}(\Phi \Phi)_{\3a}\nonumber\\
& +& i \lambda^\Phi_5 (\Phi \Phi)_{\3s}(\Phi
\Phi)_{\3a}.\\
 V(\chi) &=& \mu^2_\chi (\chi \chi)_{\s} +
\delta^\chi (\chi\chi\chi)_{\s} + \lambda^\chi_1 (\chi\chi)_{\s}
(\chi\chi)_{\s} + \lambda^\chi_2
(\chi\chi)_{\s'}(\chi\chi)_{\s''}\nonumber\\
& +& \lambda^\chi_3 (\chi\chi)_{\3}(\chi\chi)_{\3}.
\\
V(\Phi,\chi) &=& \delta^{\Phi\chi}_{s}(\Phi \Phi)_{\3s} \chi + i
\delta^{\Phi\chi}_a (\Phi \Phi)_{\3a}\chi+ \lambda^{\Phi\chi}_1
(\Phi\Phi)_{\s}(\chi\chi)_{\s}\nonumber\\
&+&\lambda^{\Phi\chi}_2 (\Phi\Phi)_{\s'}(\chi\chi)_{\s''}
+\lambda^{\Phi\chi*}_2
(\Phi\Phi)_{\s''}(\chi\chi)_{\s'}\nonumber\\
&+&\lambda^{\Phi\chi}_3 (\Phi\Phi)_{\3s}(\chi\chi)_{\3} +i
\lambda^{\Phi\chi}_4
(\Phi\Phi)_{\3a}(\chi\chi)_{\3}\, .
\label{Vacuum}
\end{eqnarray} 
Notice that the additional interactions with the Higgs field $V(H,\Phi, \chi)$ can be omitted in this context, 
since $H$ is a singlet under $A_4$ and gives rise to contributions that can be absorbed in the corresponding
coefficients of Eq.~(\ref{Vacuum}).
The self interaction terms are assumed to be confined on the two branes as in
\cite{WiseScalar}. It is then easy to check that Eq.~(\ref{eq:C3vac})
is a global minimum of $V(\Phi)$, and that
Eq.~(\ref{eq:Z2vac}) is a global minimum of $V(\chi)$.
This situation 
drastically changes once interactions between $\Phi$ and $\chi$ are
switched on via $V(\Phi,\chi)$.  The problem is that the extremal
conditions yield a larger number of independent equations than
there are unknown VEVs, as was demonstrated in \cite{Feruglio}.

To solve this problem extremely hierarchical fine tuning has to be imposed on the various parameters of the scalar potential. 
The most direct approach to avoid this fine tuning, adopted by \cite{Feruglio} and many others, was to prohibit the problematic $V(\Phi,\chi)$ interaction terms by construction.
These models are usually supersymmetric and make use of the
Froggatt-Nielsen (FN) mechanism~\cite{FN} to explain the fermion mass
hierarchy. Explaining the observed features of fermion mixings
within these setups usually requires a significantly larger flavon
content, and additional higher permutation symmetries, typically
$\mathop{\rm Z}_{n}$, in order to eliminate problematic terms from the
superpotential. The drawback is that one usually ends up with a large
parameter space without avoiding the need to make ad hoc
assumptions to account for experimental data.
  In this case, using no further assumptions, we can try to exploit the suppressed overlap of the flavons bulk profiles in order to reduce the impact of the interaction terms in Eq.~(\ref{Vacuum}) while preserving the results in the quark and lepton sectors. 
The most problematic cross-brane term is $\delta_{\Phi\chi}\Phi^2\chi$,
which is not eliminated by the additional $\mathop{\rm Z}_2$
symmetry we imposed. We know that this term is typically 
suppressed by an overlap factor  $a_{\chi}\simeq C_{\chi}/3\simeq 0.05$
compared to the self interaction term $\Phi^2$. If we switch on
the bulk mass of $\chi$, its VEV profile will become more sharply
localized on the UV brane, an effect that we can parametrize by a
multiplicative factor $C_{\mu_\chi} = e^{-\mu_{\chi}}$ in front of the original
expression in Eq.~(\ref{profiles}). This factor can easily suppress the induced shift of the
scalar VEV below the model theoretical error. The same factor enters all the calculations
performed so far, however its effect only amounts to modify the numerical value of
$C_{\chi}$ and accordingly, the matching of $c_{\nu_R}$. It turns
out that to make $a_{\chi}=(1/3)C_{\chi}C_{\mu_{\chi}}$ smaller
than the theoretical error we only need to suppress $C_{\chi}$ by a
factor of 1/2, which can be compensated by a global rescaling of
the various $x_{i}^{u,d}$ and $y_{i}^{u,d}$ parameters by at most a factor two, without
breaking the universality assumption. Quartic and higher
$V(\Phi,\chi)$ interaction terms are obviously further suppressed
and lie safely below the model theoretical error, originated by the zero mode approximation.

\subsection{Alternative solutions by model modifications}
We offer two alternative setups, in which the
desired vacuum alignment is protected by forbidding cubic mixed interaction terms, and 
 allowing only for quartic
(and higher) interactions in  $V(\Phi, \chi)$. In the first suggested setup, we use an additional
$A_4$ singlet $\eta$ and modify the external $Z_{2}$ symmetry into
$Z_{8}$. The singlet $\eta$ is in charge of the bare Majorana mass
term. Assigning the $Z_8$ charges of $\Phi$, $\chi$, $H$ and
$\eta$ to be $\alpha^4$, with $\alpha=e^{2\pi i/8}$, we assign
the rest of the fields according to
\begin{equation} (\overline{Q}_{L},
u_R', u_R'', d_R, d_R', d_R'') \Rightarrow \alpha^{4}\,, \qquad
(\overline{\ell}_{L}, \nu_R) \Rightarrow \alpha^{2}\,, \qquad
(e_R, e_R', e_R'') \Rightarrow \alpha^{6} .\label{Zeight}
\end{equation}
The above assignments allow the presence of all the operators in the Yukawa lagrangian of Eq.~(\ref{eq:Yuk}), 
while they forbid the unwanted $\overline{\ell}_{L}H\Phi\nu_R$ and
$\overline{\ell}_{L}H\chi\nu_R$ terms. Most importantly, the
dangerous cubic $\Phi^2\chi$ interaction term is also prohibited
in this setting. The operators in charge of quark mixing are now 
supplemented with an extra insertion of $\Phi$ to preserve $Z_8$, giving rise to the operators $\overline{Q}_L\Phi^2\chi u_R(u_R', u_R'', d_R, d_R', d_R'')$.
However, we have an extra source of quark mixing
allowed by the above assignments, arising from cross-brane
operators of the form
$\overline{Q}_L\chi H (u_R, u_R', u_R'', d_R, d_R', d_R'')$.\\
The latter will obviously  dominate over the contributions of
operators involving $\Phi^2\chi$ interactions. This is clearly so,
given that the above operators are of the same dimension as the
leading order operators in the Yukawa lagrangian of the quark
sector. However, being cross-brane terms, they will still be
suppressed compared to the IR dominated contributions of
$\overline{Q}_L\Phi H (u_R, u_R', u_R'', d_R, d_R', d_R'')$. After
the scale $C_\chi$ is set according to the experimental value of
$|V_{us}|$, we are able to tell if this perturbative expansion is
justified.  Due to the $Z_3$ preserving VEV of $\Phi$ and the
$Z_2$ preserving VEV of $\chi$, the above operators involving
$\chi H$ will generate the same kind of contributions as those
involving $\Phi\chi H$ to the CKM elements. However, the
corrections induced by these operators will enter in the second
row of the up and down mass matrices in the interaction basis, differently 
from Eq.~(\ref{IDCKM}). Each of these contributions will be
characterized by one coefficient which we define as $a_{\chi ,
i}^{u,d}$, where $i$ specifies the generation. When matching the
observed magnitudes of the CKM elements and the CP violating
phase, $\delta_{13}^{{\rm CKM}}$, it turns out that we still need
a relative phase $\delta_{\chi ,2}^{ud}= \pi$ between $a_{\chi
,2}^u$ and $a_{\chi ,2}^d$, to account for $V_{us}$. This results
in $C_{\chi}\simeq 0.045$, which in turn validates the
perturbative expansion, where $\epsilon_{mod.}\simeq
C_{\chi}/C_{\Phi}\simeq 0.045/0.577\simeq 0.08$ acts as the small
expansion parameter. The contributions to quark mixing arising
from the operators involving $\Phi^2\chi$  are suppressed by
$\mathcal{O}(10^{-2})$ compared to the ones associated with
$a_{\chi,i}^{u,d}$.
%%%%
It turns out that we
need two independent and non degenerate complex parameter assignments for the above
coefficients, in order to obtain an almost realistic CKM matrix at leading order as in section \ref{MatchCKM}.
This means that, in total, we have 4 non degenerate real parameters governing the
mixing data in the quark sector, a less appealing situation than the one with a $Z_2$ extra
discrete symmetry, where the two complex parameters turn out to be degenerate to a good
approximation.

The second solution we suggest is based on the simplifying
assumption that the field $\Phi$ can also play the role of the
Higgs as in many previous works on $A_{4}$
\cite{a4,Volkas,a4zee,a4ma}. It is not clear, at this stage, to
which extent this assumption can be justified in the warped setup we use,
however it seems possible
to identify the lightest mode associated with $\Phi$ in the 4D effective
theory with the SM Higgs.

We again assign a $Z_8$ discrete symmetry in this slightly
simplified setup. Given that $\Phi$, $\chi$, and $\eta$  transform
again as $\alpha^4$, we choose the other fields to transform
according to
\begin{equation} (\overline{Q}_{L})
\Rightarrow \alpha^{4}\,, \qquad (\nu_R) \Rightarrow \alpha^{2}\,,
\qquad (\ell_L, \phi) \Rightarrow \alpha^3\,, \qquad (e_R, e_R',
e_R'') \Rightarrow \alpha \, ,\label{Zeight2}
\end{equation}
and importantly the scalar sector is now  supplemented with an
additional $A_4$ singlet $\phi$ which is in charge of the neutrino
Dirac mass term. We naturally expect $\eta$ and $\phi$ to be UV
and IR localized, respectively. All terms in the Yukawa lagrangian
of Eq.~(\ref{eq:Yuk}) are still allowed by the above assignment
with $H$ being swallowed into $\Phi$, and the Dirac mass term for the
neutrinos of the form $\overline{\ell}_L\phi\nu_R$. The
dominant higher order corrections in the Dirac neutrino mass
matrix arise from operators of the form
$\overline{\ell}_L\phi\Phi^2 \nu_R$ and
$\overline{\ell}_L\phi\chi^2 \nu_R$, for which the associated
contributions were already inspected in section III. The dominant
higher order corrections to the heavy Majorana mass matrix will
consist of $\chi^3\nu_R(\nu_R)^c$ and $\Phi^2\eta\nu_R\nu_R^c$,
for which the resulting textures were also inspected in the same
section.

Turning back to the quark sector we see that the most dominant
cross-talk interactions, leading to quark mixing, are of the form
$\overline{Q}_L\Phi^2\chi q^{u,d}_R$ and $\overline{Q}_L\Phi\chi^2
q^{u,d}_R$, both of which give similar contributions to those
described in section~\ref{Crosstalk1}. Consequently, we can match
the CKM matrix as we already did in section~\ref{FindCKM}.
Differences will stem from redefinitions of  $\Phi_0$ and $C_\chi$
and consequent rescaling of fermion bulk masses. The resulting
mixing data in the quark sector will be now governed by
$\tilde{x}_i^{u,d}$, $\tilde{y}_i^{u,d}$ and $\tilde{z}_i^{u,d}$,
with the newly defined $\tilde{z}_i^{u,d}$ parameters entering at
the second row of the up and down mass matrices in the interaction
basis. A matching of an almost realistic CKM matrix by 4 real
parameter assignments analogous to the one in section
\ref{FindCKM} can be performed, yet the parameter space is still
larger. As already said another possible drawback of this proposal
may lie in the identification of the 5D flavon $\Phi$ with the
bulk Higgs.

We should also add that
in both solutions offered above to protect the desired
vacuum alignment, the  $G_{\rm SM}^{\rm cust}\times A_4\times Z_8$
invariant interaction terms
in the scalar potential with insertions of the new fields $\eta$
and $\phi$ are irrelevant to the vacuum alignment problem,
since they are both  flavor singlets. Further constructions with
additional fields and more complicated flavor symmetries are
obviously possible at the price of an increased arbitrariness of the model.

\section{Flavor violation and the Kaluza-Klein scale}
\label{sec:FV}
 All models with extra dimensions will have to face the presence of mixing between the
degrees of freedom of the effective 4D theory and their KK excitations. 
One of the crucial tasks in the construction of these models, is thus to guarantee that 
 corrections induced by this mixing do not spoil the agreement with observations for a
 natural value of the lowest KK scale of order a few TeV. 
In other words, once new physics (NP) contributions
induced by the exchange of the KK excitations are taken into account, 
the agreement with observations 
will force a lower bound on the KK scale, and we demand it be naturally of order a few TeV.

It has been shown \cite{Agashe:2003zs} that custodial symmetry in the bulk of RS warped 
models is able to reduce the lower 
bound on the first KK mass imposed by electroweak precision measurements from $\sim 10$ TeV to 
$\sim 4$ TeV. A generalization of that analysis to a wider set of scenarios and including higher order corrections can be found in \cite{Carena:2007}.
More recently, it has been observed \cite{Agashe:2004cp, Azatov, IsidoriPLB} that FCNC 
processes can in general produce more stringent bounds than the observed S, T parameters 
on the KK scale, and that a residual CP problem remains in the form of excessive contributions to $\epsilon_K$ \cite{Agashe:2004cp}, the direct CP violation parameter $\epsilon^\prime/\epsilon_K$ \cite{IsidoriPLB} and the neutron electric dipole moment (EDM) \cite{Agashe:2004cp}.  

These bounds can slightly be improved if the Higgs field is allowed to propagate in the bulk. In this case all zero mode fermions can be pushed further towards the UV brane, preserving the same 4D mass and having a reduced overlap with the IR localized KK modes. As a consequence
 the 5D Yukawa couplings can be raised without violating perturbativity constraints.  
As we already observed, this is particularly relevant for the top quark, being it the
 heaviest fermion zero mode and the most IR localized. 
For these reasons, our model realizes custodial symmetry with a bulk Higgs, in addition to an $A_4$ discrete bulk flavor symmetry.

The predictions and constraints derived in \cite{Agashe:2004cp,IsidoriPLB} apply to the 
general case of flavor anarchical 5D Yukawa couplings. The conclusions may differ
 if a flavor pattern of the Yukawa couplings is assumed to hold in the 5D theory due to bulk flavor symmetries. They typically imply an increased alignment between the 4D fermion mass 
matrix and the Yukawa and gauge couplings, thus suppressing the amount of flavor violation 
induced by the interactions with KK states.
In our case, the most relevant consequence of the $A_4$ flavor symmetry is  the degeneracy of the left-handed fermion bulk profiles $f_Q$, i.e. $diag(f_{Q_1,Q_2,Q_3})=f_Q\times \mathbbm{1}$. In addition, the distribution of phases, CKM and Majorana-like, in the mixing matrices
might induce zeros in the imaginary components of the Wilson coefficients contributing to CP violating quantities.
%%%%%%%%%%%%%%%%%%%%%%%%%%%%%%%%%%%%%%%%%%%%%%%%%%%%%
%% 
\begin{center}
\begin{eqnarray}\nonumber
\parbox{35mm}{\vspace*{-1.5cm}
\begin{fmfgraph*}(30,30)
\fmfleft{p1,p2}
\fmfright{p3,p4}
\fmf{fermion,label=${\scriptstyle d}$}{p1,v1}
\fmf{fermion,label=${\scriptstyle \bar s}$}{v1,p2}
\fmf{gluon,label=${\scriptstyle G^{(n)^{ }}}$,label.side=right,label.dist=3mm}{v1,v2}
\fmf{fermion,label=${\scriptstyle s}$}{v2,p3}
\fmf{fermion,label=${\scriptstyle \bar d}$}{p4,v2}
\end{fmfgraph*}  }
\parbox{60mm}{
\begin{fmfgraph*}(50,50)
\fmfleft{p1}
\fmfright{p2}
\fmftop{p3}
\fmf{fermion,tension=1.7,label=${\scriptstyle d_R(s_R)}$,label.side=right}{v9,p2}
\fmf{plain,tension=4.}{v2,v9}
\fmf{fermion,tension=1.7,label=${\scriptstyle d_L(b_L)}$,label.side=right}{p1,v10}
\fmf{plain,tension=4.}{v10,v1}
\fmf{photon,tension1.5,label=${\scriptstyle \gamma}$}{p3,v3}
\fmf{dashes,tension=.8,label=${\scriptstyle {H}}$,label.side=right}{v1,v2}
\fmf{fermion,tension=.8}{v1,v45}
\fmf{plain,tension=5,label=${\scriptstyle d^{(n)}_R}$,label.side=left}{v45,v4}
\fmf{crossed,tension=1}{v4,v5}
\fmf{fermion,tension=.7,label=${\scriptstyle d^{(l)}_L}$,label.side=left}{v5,v3}
\fmf{plain,tension=.7}{v3,v6}
\fmf{fermion,tension=1,label=${\scriptstyle d^{(k)}_L}$,label.side=left}{v6,v7}
\fmf{plain,tension=.8}{v7,v75}
\fmf{plain,tension=5}{v75,v2}
\end{fmfgraph*}  }
\parbox{60mm}{
\begin{fmfgraph*}(50,50)
\fmfleft{p1}
\fmfright{p2}
\fmftop{p3}
\fmf{fermion,tension=1.7,label=${\scriptstyle d_R(s_R)}$,label.side=right}{v9,p2}
\fmf{plain,tension=4.}{v2,v9}
\fmf{fermion,tension=1.7,label=${\scriptstyle d_L(b_L)}$,label.side=right}{p1,v10}
\fmf{plain,tension=4.}{v10,v1}
\fmf{photon,tension1.5,label=${\scriptstyle \gamma}$}{p3,v3}
\fmf{dashes,tension=.8,label=${\scriptstyle {H^\pm}}$,label.side=right}{v1,v2}
\fmf{fermion,tension=.8}{v1,v45}
\fmf{plain,tension=5,label=${\scriptstyle u^{(n)}_R}$,label.side=left}{v45,v4}
\fmf{crossed,tension=1}{v4,v5}
\fmf{fermion,tension=.7,label=${\scriptstyle u^{(l)}_L}$,label.side=left}{v5,v3}
\fmf{plain,tension=.7}{v3,v6}
\fmf{fermion,tension=1,label=${\scriptstyle u^{(k)}_L}$,label.side=left}{v6,v7}
\fmf{plain,tension=.8}{v7,v75}
\fmf{plain,tension=5}{v75,v2}
\end{fmfgraph*} 
}
\end{eqnarray}
\end{center}
{\small Fig. 2: Contribution to $\epsilon_K$ from a KK gluon exchange (left). Dipole contributions to the neutron EDM from KK down-quarks (center) and KK up-quarks (right). The same type of dipole diagrams also contribute to $b\to s(d)\gamma$ and $\epsilon^\prime/\epsilon_K$.} 
\vspace{0.8truecm}
%%%%
We can straightforwardly identify a few properties of our model and state some results, while we defer to a separate work a more complete study. 
Following the spurion analysis in \cite{Agashe:2004cp}, we observe the following facts, 
consequence of the degenerate left-handed fermion profiles. First, the new physics contribution to $\epsilon_K$ coming from a KK gluon exchange, the leftmost diagram in 
fig.~2 vanishes, since
\begin{equation}
 \epsilon_K^{NP}\propto {\mbox Im} \left [ V_L^{d\dagger}{\mbox diag}(f^{-2}_{Q_1,Q_2,Q_3}) V_L^d \right ]_{12}^2 =
f_Q^{-4}\, {\mbox Im} ( V_L^{d\dagger}V_L^d)_{12}^2  =0\, ,
\end{equation}
in the basis where the fermion profiles are real and diagonal. We notice that the amplitude itself vanishes and not only its imaginary part. Second, and for the same
 reason, the left-handed rotation matrix $V_L^d$ will disappear from all down-type 
contributions to the dipole effective operators of the form 
$O_{\gamma ,(g)}^{ij} = \bar{d}_L^i\sigma_{\mu\nu} d_R^j F(G)^{\mu\nu}$, center diagram in 
 Fig.~1. A pleasing result is the vanishing of the down-type new physics 
contributions to the neutron electric dipole moment, induced by the same diagram 
with external $d$ quarks. 
Again following \cite{Agashe:2004cp}, this contribution vanishes as follows 
\begin{eqnarray}
{\mbox Im} C_{d_n}^{d-type} &\propto& 
{\mbox Im} \left [ V_R^{d\dagger}{\mbox diag}(f^{2}_{d_1,d_2,d_3}) V_R^d{\mbox diag} (m_{d,s,b})
 V_L^{d\dagger}{\mbox diag}(f^{2}_{Q_1,Q_2,Q_3}) V_L^d \right ]_{11} \nonumber\\
&=&f_Q^2 m_d {\mbox Im} \left [ V_R^{d\dagger}{\mbox diag}(f^{2}_{d_1,d_2,d_3}) V_R^d\right ]_{11}   =0\, .
\end{eqnarray}
Contributions from up-type KK fermion exchange, rightmost diagram in fig.~2,
involve left- and right-handed matrices and are thus  expected to be non zero, generally 
of the same size as in any flavor anarchic model. However, a bulk flavor symmetry might 
induce an interesting cancellation of observable phases, so that dominant new physics 
contributions to the neutron electric dipole moment and, or, to 
$\epsilon^\prime /\epsilon_K$ will vanish. We leave this analysis and the study of Higgs 
mediated FCNCs for future work. 
We conclude here that the presence of a bulk $A_4$ flavor symmetry can only improve upon the residual CP violation problem that in general affects warped models with flavor anarchy.
 We have seen that the constraint induced by tree level KK gluon exchange to $\epsilon_K$
 is released, leaving $b\to s(d)\gamma$, $\epsilon^\prime /\epsilon_K$, 
the neutron EDM and Higgs mediated FCNCs as possible candidates to produce the most 
stringent 
lower bounds on the KK scale. In the worst scenario, a milder lower bound from the
EDM and $\epsilon^\prime /\epsilon_K$ is expected in our model due to the vanishing of 
down-type dipole contributions. On the other hand, the degeneracy of left-handed fermion bulk profiles is expected to not be 
sufficient to produce a suppression of 
Higgs mediated FCNCs at tree level. This might require additional constraints on the right-handed fermion profiles. 

Alternatively, a cancellation of observable phases, and a consequent vanishing of the 
imaginary parts of flavor violating amplitudes such as the new physics contributions to 
the EDM, $\epsilon_K$ and  $\epsilon^\prime /\epsilon_K$, would obviously be a welcomed 
feature of the model; it would release the most stringent lower 
bounds on the KK scale from CP violating observables and solve the residual CP problem.

\section{Conclusions}
\label{sec:conclusion}

We have constructed a warped extradimensional realization of an $A_4$ flavor model for 
quarks and leptons, and implemented the flavor symmetry breaking pattern 
$A_4\to {\rm nothing}$ first suggested in \cite{Volkas}.
In this construction all standard model fields, including the Higgs field, propagate in 
the bulk and a bulk custodial symmetry
is broken in two different ways \cite{Agashe:2003zs} on the UV and IR brane by orbifold boundary conditions.
The spontaneous symmetry breaking of the $A_4$ flavor symmetry is induced by the VEVs of 
two bulk flavon fields $\Phi$ and $\chi$: $\Phi$ is responsible of the breaking pattern 
$A_4\to Z_3$ in the charged fermion sector, while $\chi$ is responsible of the breaking 
pattern $A_4\to Z_2$ in the neutrino sector. By taking the two flavons to be peaked on 
different branes, 
we approximately sequester the two sectors and the associated symmetry breaking patterns: 
neutrinos with the UV-peaked $\chi$ on one side, charged fermions 
with the IR-peaked $\Phi$ on the other. 
If the two sectors do not communicate, that is when the interactions of $\Phi$ with neutrinos and $\chi$ with charged fermions are switched off, tribimaximal mixing for neutrinos is 
exactly reproduced, while no quark mixing is generated.
In our model the two flavons propagating in the bulk are responsible for cross-brane 
interactions and a complete cross-talk between the charged fermion and neutrino sectors. As a consequence, quark mixing on the IR brane is generated by contributions which are naturally suppressed by the warped geometry with respect to the leading order pattern in the quark and lepton sectors \cite{a4,Volkas}.

Using this realization we have obtained an almost realistic CKM matrix, including its CP violating phase, with almost degenerate order one complex Yukawa couplings. The 
large hierarchy of standard model fermion masses is generated by a tiny hierarchy in
 the bulk fermion mass parameters, a well known pleasing feature of warped constructions.
At the same time the contributions of all cross-talk and cross-brane effects do not spoil the tribimaximal mixing pattern in the neutrino sector, where they produce 
deviations within 
$1\sigma$ from the experimental values, with small non zero contributions also to 
$\theta_{13}$.

The cross-talk/brane induced quark mixing, to leading order in the perturbative
diagonalization of the mass matrices, is expressed in terms of six complex parameters in the up and down sectors, 
$\tilde{x}_{i}^{u,d},\, \tilde{y}_{i}^{u,d}$, with $i=1,2,3$. 
It turns out that, with all these parameters of order one and allowing for at least two relative phases, one obtains an almost realistic CKM matrix within the model theoretical error. 
At this order the diagonal CKM entries remain equal to one, and the two smallest entries 
$V_{ub}$ and $V_{td}$ are degenerate. 

We have also noticed that the possibility of producing hierarchical CKM matrix elements, of order 
$\lambda_{CKM}$, $\lambda_{CKM}^2$ and  $\lambda_{CKM}^3$, with all parameters of order one 
stems from the 
presence of built-in cancellations induced by the hierarchical masses. 
The presence of the $A_4$ induced phase $\omega$ also produces a pattern in the corrections.
Analogous cancellations are expected to occur at higher orders.
It is instructive to compare this $A_4$ pattern with other flavor symmetry groups, in 
particular $T^\prime$ \cite{TPrime}. The Wolfenstein parametrization would 
suggest the 
existence of an expansion parameter to be naturally identified with $\lambda_{CKM}$, 
offering an elegant and simple description of quark mixing in the standard model.
On the other hand, flavor models based on otherwise appealing discrete flavor symmetries 
such as $A_4$ and $T^\prime$ seem to rely on more complicated patterns in order to produce a realistic 
CKM matrix. 

A bulk $A_4$ flavor symmetry is also  welcomed in order to suppress the amount of flavor 
violation induced by the mixing of the standard model particles -- the zero modes
of the 5D theory -- with their Kaluza-Klein excitations. The degeneracy of left-handed 
fermion bulk profiles, due to having assigned the left-handed fermions to triplets of $A_4$, 
 implies that the tree level contribution from a KK gluon exchange to $\epsilon_K$ vanishes. 
For the same reason, all down-type dipole contributions to the neutron electric dipole 
moment and $\epsilon^\prime /\epsilon_K$ also vanish. 
The situation is different for Higgs mediated FCNCs \cite{Azatov} and their contribution to $\epsilon_K$ at tree level, since they involve both left- and right-handed fermion profiles. Their suppression might require further constraints on the right-handed sector, to be explored
in future work.
Even in the presence of non vanishing amplitudes, an $A_4$ induced cancellation of observable
 phases and the consequent vanishing of new physics contributions to the EDM, 
$\epsilon_K$ and  $\epsilon^\prime /\epsilon_K$, would obviously be a welcomed feature of 
the model, removing the most stringent bounds on the KK scale and resolving the little CP
 problem \cite{Agashe:2004cp}.

Finally, the presence of cross-brane interactions of the flavon fields $\Phi$ and $\chi$ inevitably induces 
deviations from the VEVs that realize the two breaking patterns $A_4 \to Z_3$ and $A_4\to Z_2$, leading to the well known vacuum alignment problem.  
However, in this case such corrections are naturally suppressed, being the two flavons 
peaked on different branes. In particular, 
we have shown that the contribution from the most dominant term in the interaction potential $V(\Phi, \chi)$ can be pushed below the model theoretical error by introducing a bulk mass for the UV-peaked $\chi$ field.
Obviously, this implies the need to rescale by a global amount all the $\tilde{x}_{i}^{u,d}$, $\tilde{y}_{i}^{u,d}$ parameters 
entering quark mixing, a rescaling that should anyway maintain all parameters of order one and satisfy all perturbativity bounds.
On the other hand, employing a $Z_{8}$ symmetry setup, as suggested in section VI, 
directly forbids the most dangerous terms and seems to provide a more elegant solution.

The $A_{4}$ flavor symmetry still appears to be the most elegant and economical way
to account for the nearly tribimaximal mixing pattern of neutrinos.
We have shown here that it is also possible to obtain an almost realistic quark
mixing, using a rather simple embedding of $A_4$ in a warped extra dimension and with minimal field content.
The main advantage of this construction remains the one of having cross-brane interactions and cross-talk
effects sufficiently large to account
for the observed quark mixing, without affecting the other results, or spoiling the vacuum alignment. 
A dynamical completion of this, as of other flavor models involving a discrete flavor symmetry would certainly be desirable. 
Possible
scenarios already described in the literature include a spontaneous symmetry breaking of a
continuous flavor symmetry \cite{GrossNew} or having $A_4$ as a
remnant spacetime symmetry of a toroidal compactification scheme
of a six-dimensional spacetime \cite{Altars}.

It is worth to mention two additional points. It is appealing to explore the effects of a heavier Higgs in this context. Some of the model predictions would get closer to the best fits for certain observables, such as $Zb\bar{b}$ ratios and asymmetries, allowing for a larger parameter space. 
However, as discussed in the literature, a heavier Higgs mass in a custodial setup easily induces conflicts with electroweak precision measurements and a more careful estimate of the allowed range of values for $m_H$ should be produced in each version of the custodial setup.
The final point concerns possible extensions of the warped $A_4$ model. 
It is interesting and phenomenologically relevant to consider the possibility to embed $P_{LR}$ (or other versions of) custodial symmetry \cite{Agashe:2006at} into warped $A_4$. This would release the most stringent constraint on the model parameter space due to the $Zb\bar{b}$ best fits and could provide new appealing features alternative to flavor anarchic models.

\vspace*{1.0truecm}
{\bf \noindent Acknowledgments}
%\acknowledgments  
\vspace*{0.2truecm}

\noindent We thank Yuval Grossman and Gilad Perez for useful discussions.
The work of A.K is supported in part by the Ubbo Emmius scholarship program at the University of Groningen.

\appendix

\section{Basic $A_4$ properties.}
The alternating group of order four, denoted $A_4$, is defined as
the set of all twelve even permutations of four objects and is
isomorphic to $T$, the tetrahedral group. It has a real
three-dimensional irreducible representation $\3$, and three
inequivalent one-dimensional representations $\s$, $\spr$ and
$\sppr$.  The representation $\s$ is trivial, while $\spr$ and
$\sppr$ are non-trivial and complex conjugates of each other.

The twelve representation matrices for $\3$ are conveniently taken
to be the $3 \times 3$ identity matrix $1$, the reflection
matrices $r_1 \equiv {\rm diag}(1,-1,-1)$, $r_2 \equiv {\rm
diag}(-1,1,-1)$ and $r_3 \equiv {\rm diag}(-1,-1,1)$, the cyclic
and anticyclic matrices
\begin{equation}
c = a^{-1} \equiv \left( \begin{array}{ccc} 0 & 0 & 1 \\ 1 & 0 & 0
\\ 0 & 1 & 0 \end{array} \right)\quad {\rm and}\quad a = c^{-1}
\equiv \left( \begin{array}{ccc} 0 & 1 & 0 \\ 0 & 0 & 1 \\ 1 & 0 &
0 \end{array} \right), \label{eq:ca}
\end{equation}
respectively, as well as $r_i c r_i$ and $r_i a r_i$.  Under the
group element corresponding to $c (a)$, $\spr \to \omega
(\omega^2) \spr$ and $\sppr \to \omega^2 (\omega) \sppr$, where
$\omega = e^{i2\pi/3}$ is a complex cube root of unity, with both representations 
being unchanged under the $r_i$.

The basic non-trivial tensor products are
\begin{equation}
\3 \otimes \3 = \3_s \oplus \3_a \oplus \s \oplus \spr \oplus
\sppr,\quad {\rm and}\quad \spr \otimes \spr = \sppr,
\label{eq:A4tensors}
\end{equation}
where $s (a)$ denotes symmetric (antisymmetric) product.  Let
$(x_1,x_2,x_3)$ and $(y_1,y_2,y_3)$ denote the basis vectors for
two $\3$'s. Then
\begin{eqnarray}
(\3 \otimes \3)_{\3s}  & = & ( x_2 y_3 + x_3 y_2\, ,\, x_3 y_1 +
x_1 y_3\, ,\, x_1 y_2 + x_2 y_1 ),
\label{eq:33to3s}\\
(\3 \otimes \3)_{\3a}  & = & ( x_2 y_3 - x_3 y_2\, ,\, x_3 y_1 -
x_1 y_3\, ,\, x_1 y_2 - x_2 y_1 ),
\label{eq:33to3a}\\
(\3 \otimes \3)_{\s} & = & x_1 y_1 + x_2 y_2 + x_3 y_3, \label{eq:33tos}\\
(\3 \otimes \3)_{\spr} & = & x_1 y_1 + \omega\, x_2 y_2 + \omega^2\, x_3 y_3, \label{eq:33tospr}\\
(\3 \otimes \3)_{\sppr} & = & x_1 y_1 + \omega^2\, x_2 y_2 +
\omega\, x_3 y_3, \label{eq:33tosppr}
\end{eqnarray}
in an obvious notation.

\end{fmffile}
\end{document}